\documentclass[usenatbib]{mnras}

\usepackage{newtxtext,newtxmath}

\usepackage[T1]{fontenc}
\usepackage{ae,aecompl}

\usepackage{graphicx}	
\usepackage{amsmath}	
\usepackage{subcaption}
\usepackage{placeins}

\usepackage[capitalise,noabbrev]{cleveref} 

\graphicspath{{./figs/}}
\DeclareGraphicsExtensions{.pdf,.png}

\newcommand{\kpc}{\,\textrm{kpc}} 
\newcommand{\myr}{\,\textrm{Myr}} 
\newcommand{\rcore}{r_\textrm{c}} 

\renewcommand{\vec}[1]{\boldsymbol{#1}}

\defcitealias{Yates-Jones2021}{Paper~I}
\defcitealias{Turner2018a}{RAiSE~II}

\title[PRAiSE]{PRAiSE: Resolved spectral evolution in simulated radio sources}

\author[Patrick M. Yates-Jones et al.]{%
Patrick M. Yates-Jones,$^{1,2,3}$\thanks{E-mail: patrick.yates@utas.edu.au}
Ross J. Turner,$^{1}$
Stanislav S. Shabala,$^{1,2,3}$\newauthor
and Martin G. H. Krause$^{2}$
\\
$^{1}$ School of Natural Sciences, University of Tasmania, Private Bag 37, Hobart, TAS 7001, Australia\\
$^{2}$ Centre for Astrophysics Research, University of Hertfordshire, College Lane, Hatfield, Herts AL10 9AB, UK \\
$^{3}$ ARC Centre of Excellence for All Sky Astrophysics in 3 Dimensions (ASTRO 3D), Australia\\
}

\date{Accepted XXX. Received YYY; in original form ZZZ}

\pubyear{2021}

\begin{document}
\label{firstpage}
\pagerange{\pageref{firstpage}--\pageref{lastpage}}
\maketitle

\begin{abstract}
  We present a method for applying spatially resolved adiabatic and radiative loss processes to synthetic radio emission from hydrodynamic simulations of radio sources from active galactic nuclei (AGN).
  Lagrangian tracer particles, each representing an ensemble of electrons, are injected into simulations and the position, grid pressure, and time since the last strong shock are recorded.
  These quantities are used to track the losses of the electron packet through the radio source in a manner similar to the \textit{Radio AGN in Semi-analytic Environments} (RAiSE) formalism, which uses global source properties to calculate the emissivity of each particle ex-situ.
  Freedom in the choice of observing parameters, including redshift, is provided through the post-processing nature of this approach.
  We apply this framework to simulations of jets in different environments, including asymmetric ones.
  We find a strong dependence of radio source properties on frequency and redshift, in good agreement with observations and previous modelling work.
  There is a strong evolution of radio spectra with redshift due to the more prominent inverse-Compton losses at high redshift.
  Radio sources in denser environments have flatter spectral indices, suggesting that spectral index asymmetry may be a useful environment tracer.
  We simulate intermediate Mach number jets that disrupt before reaching the tip of the lobe, and find that these retain an edge-brightened Fanaroff-Riley Type II morphology, with the most prominent emission remaining near the tip of the lobes for all environments and redshifts we study.
\end{abstract}





\section{Introduction}

  Synchrotron emission from high-energy electrons with a non-thermal energy distribution is ubiquitous both within the Milky Way galaxy \citep[e.g.,][]{Westerhout1958,Higdon1979,Haslam1981,Jaffe2013,Carretti2013,Green2019,Becker2021} and from extragalactic radio sources \citep[e.g.,][]{Baade1956,Burbidge1956,Perley1982,Heesen2015,Krause2021}.
  Extragalactic radio jets are capable of accelerating particles to high energies \citep{Matthews2020}, followed by losses due to several mechanisms.
  The dynamics of radio sources play a significant role in the observed emission, as adiabatic, synchrotron, and inverse-Compton loss processes depend on the cocoon dynamics.
  The populations of electrons accelerated at strong shocks both in the jet and at terminal shocks will flow into the radio lobes, often with substantial mixing \citep{Turner2018a}.
  Flow dynamics therefore play a role in shaping the spatially-resolved lobe radio spectra.
  Accounting for both this mixing of electron populations and the flow dynamics is crucial to interpreting observations of radio lobes.

  Analytic and semi-analytic models exist \citep{Scheuer1974,Begelman1989,Falle1991,Kaiser1997,Kaiser1997a,Blundell2000,Manolakou2002,Luo2010,Shabala2013a,Maciel2014,Turner2015,Turner2018a,Hardcastle2018} to model both the dynamics and emissivity of radio sources, including both adiabatic and radiative losses.
  However, only hydrodynamic simulations are able to fully capture the complex turbulence and mixing that occurs in real sources.
  Numerical simulations can also better model the magnetic field, which plays a key role in the synchrotron energy loss process.
  \citet{Hardcastle2014} showed using magnetohydrodynamic simulations that the magnetic field energy density varies spatially within a pair of radio lobes.
  This introduces a spatial dependence into the radiative loss process and further emphasises the need for simulations.

  Several efforts have been made over the years to include particle acceleration and losses in numerical fluid simulations.
  Both electron transport using tracer fluids \citep{Jones1999,Tregillis2001,Tregillis2004} and non-thermal test particles \citep{Mimica2009,Obergaulinger2015,Fromm2016,Fromm2018,Fromm2019} have been used in (relativistic) hydrodynamic simulations.
  Recently, this approach has been extended to incorporate the effect of magnetic fields \citep{Mendygral2012,Vaidya2018,Mukherjee2020} to self-consistently accelerate particles with diffusive shock acceleration.
  \citet{Walg2020} explicitly model the electrons using a two-fluid approach.

  A challenge for all these approaches is the computational cost associated with each simulation.
  Realistic radio galaxy simulations require low jet densities and high velocities, thus demanding many small computational steps \citep[e.g.,][]{Krause2003,Krause2005}.
  Unless a minimum, code-dependent numerical resolution is maintained, vortex shedding at the jet head and interaction of turbulent vortices in the cocoon with the jet beam will not be captured correctly and the source expansion will be too fast \citep{Krause2001}.
  Instabilities in the jet, which may have important consequences for radio source morphology, can also only be captured with high-resolution simulations \citep{Massaglia2016}.

  The literature methods discussed above evolve electron packets in-situ, according to the local histories of pressure, magnetic and radiation fields they experience on their way through the radio sources.
  Observing a simulated source at a different redshift, or varying the strength of the radiation field of the host galaxy requires re-running the whole simulation.

  As an alternative to the tracer fluid approach, Lagrangian tracer particles can be used to capture dynamical information about the simulation state for subsequent ex-situ analysis \citep[e.g.,][]{Harlow1965}.
  This method has been applied to simulations in order to follow thermodynamic quantities to trace accretion in galaxies \citep{Genel2013}, study acceleration of cosmic ray protons \citep{Wittor2016,Wittor2017,Vazza2021}, and gain insight into AGN-driven turbulence \citep{Wittor2020}.

  In this paper we present a hybrid approach to modelling radio source evolution, combining a grid-based fluid simulation framework with Lagrangian tracer particles.
  The Particles+RAiSE (PRAiSE) framework uses Lagrangian tracer particles that are evolved with the simulation, and records the histories of pressure, magnetic field, and shock passages.
  In post-processing, an observing frequency is chosen for each particle at a given snapshot time, and the radiation field is chosen.
  The emitting particle Lorentz factor is then interpolated backwards to the last acceleration event, and the synchrotron emissivity is calculated accordingly.
  In this way, we can use the same hydrodynamic simulation to produce synthetic radio images at different redshifts, and using different assumptions about injection energy distributions of the radiating particles.

  We summarise the details of the PRAiSE implementation in \cref{sec:sr_praise_implementation}.
  In \cref{sec:sr_validation}, the feasibility of this method is demonstrated with a high-resolution jet simulation that includes an unstable jet similar to the simulations in \citet{Massaglia2016}.
  We find that particle acceleration at localised shock regions even in the unstable jet are captured well.
  We apply the code to the case of radio sources in asymmetric environments in \cref{sec:sr_results}, showing that the spectral index correlates with the local environment, and conclude in \cref{sec:sr_discussion} with a discussion of our method and results.

  The spectral index $\alpha$ is defined by $S_\nu=\nu^{-\alpha}$ for flux density $S$ and frequency $\nu$ throughout this paper.

\section{PRAiSE implementation}
\label{sec:sr_praise_implementation}

  \begin{figure*}
    \centering
    \includegraphics{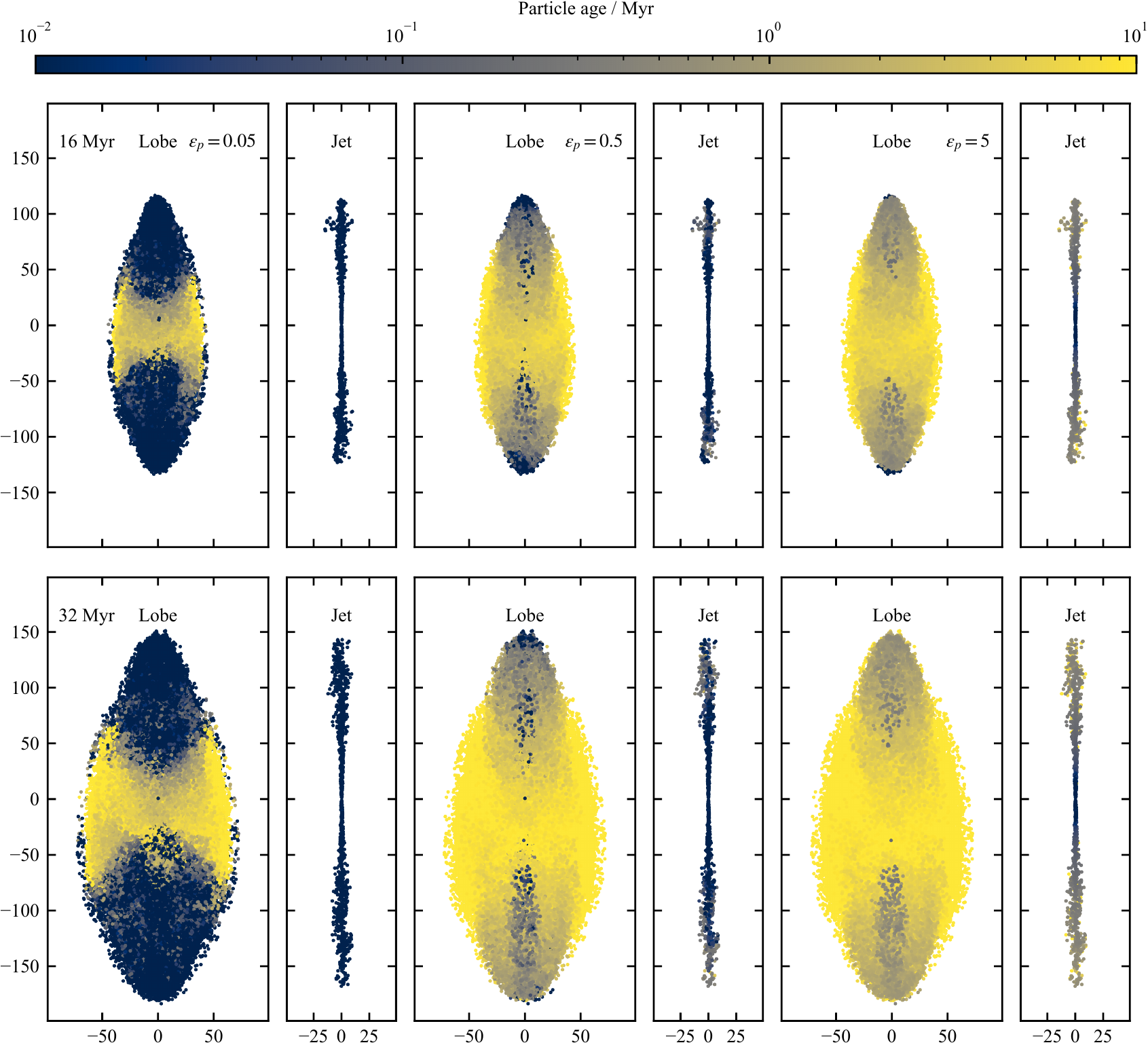}
    \caption{
      Particle ages for the $0\rcore$-offset simulation, at $t=16\myr$ (upper panels) and $t=32\myr$ (lower panels), and for different pressure thresholds used for shock detection, $\epsilon_p=0.05, 0.5, 5.0$ (from left to right).
      Particles belonging to either the lobes or jet are separated using a velocity cut of $|\textrm{v}|=0.3c$ as described \cref{sec:sr_spatially_resolved_losses}.
      These populations are plotted side-by-side for a given pressure threshold (only labelled on the top lobe panels).
    }
    \label{fig:sr_shock_threshold}
  \end{figure*}

  \begin{figure*}
    \centering
    \includegraphics{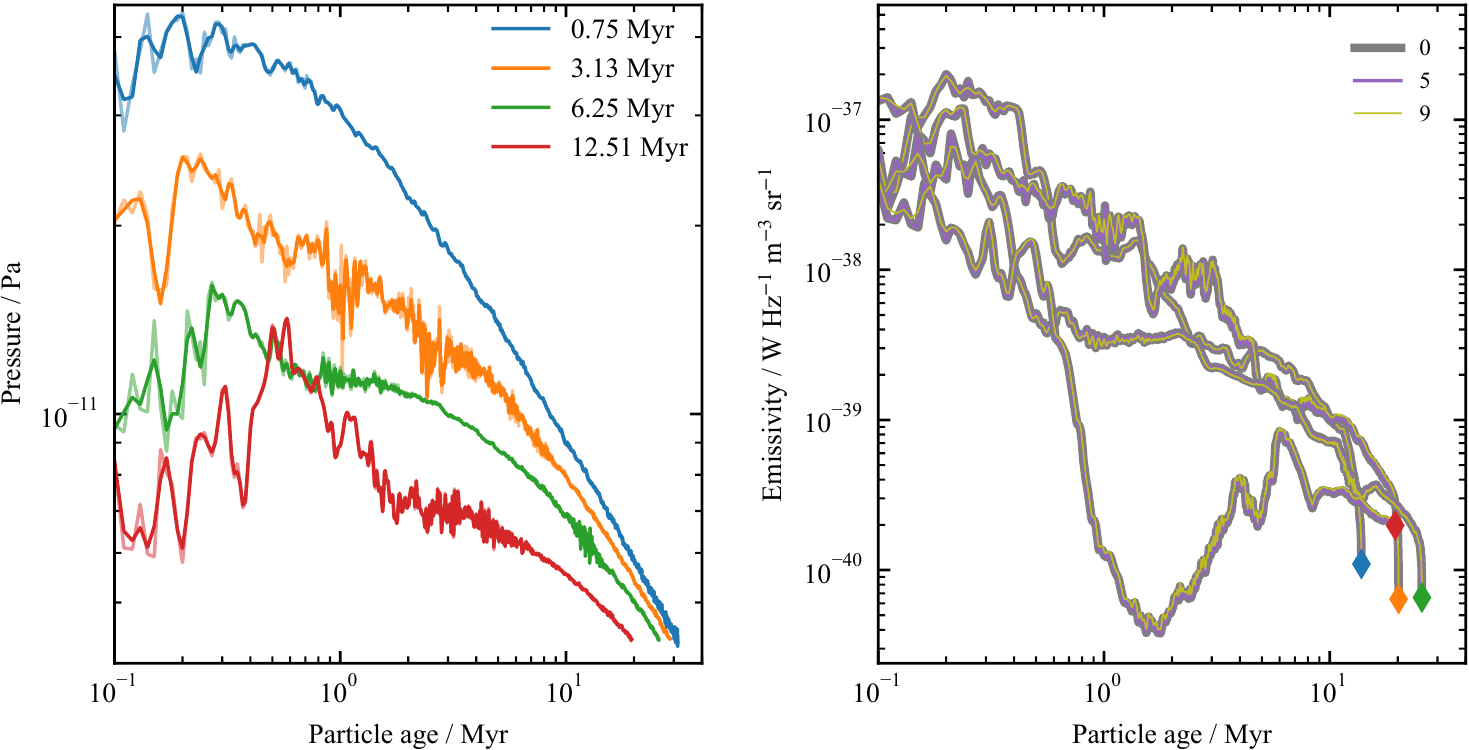}
    \caption{
      \textit{Left}: Pressure evolution as a function of time for four randomly selected Lagrangian tracer particles.
      Semi-transparent lines show the unsmoothed pressure, while the opaque lines are the smoothed pressure.
      A Savitzky--Golay filter with window length $5$ and polynomial order $3$ is used.
      The abscissa is the particle age since it was last shocked.
      The pressure tracks are labelled by the most recent shock time.
      \textit{Right}: Emissivity evolution as a function of time, for the same particles as the left-hand panel.
      Emissivity tracks for two different pressure smoothing window lengths are shown ($5$, $9$), in addition to emissivity calculated using unsmoothed data (labelled $0$).
      The diamonds mark the particle age when the unsmoothed emissivity drops to zero; the diamond colours correspond to the individual particles in the left-hand panel.
    }
    \label{fig:sr_particle_smoothing}
  \end{figure*}

  \begin{figure*}
    \centering
    \includegraphics{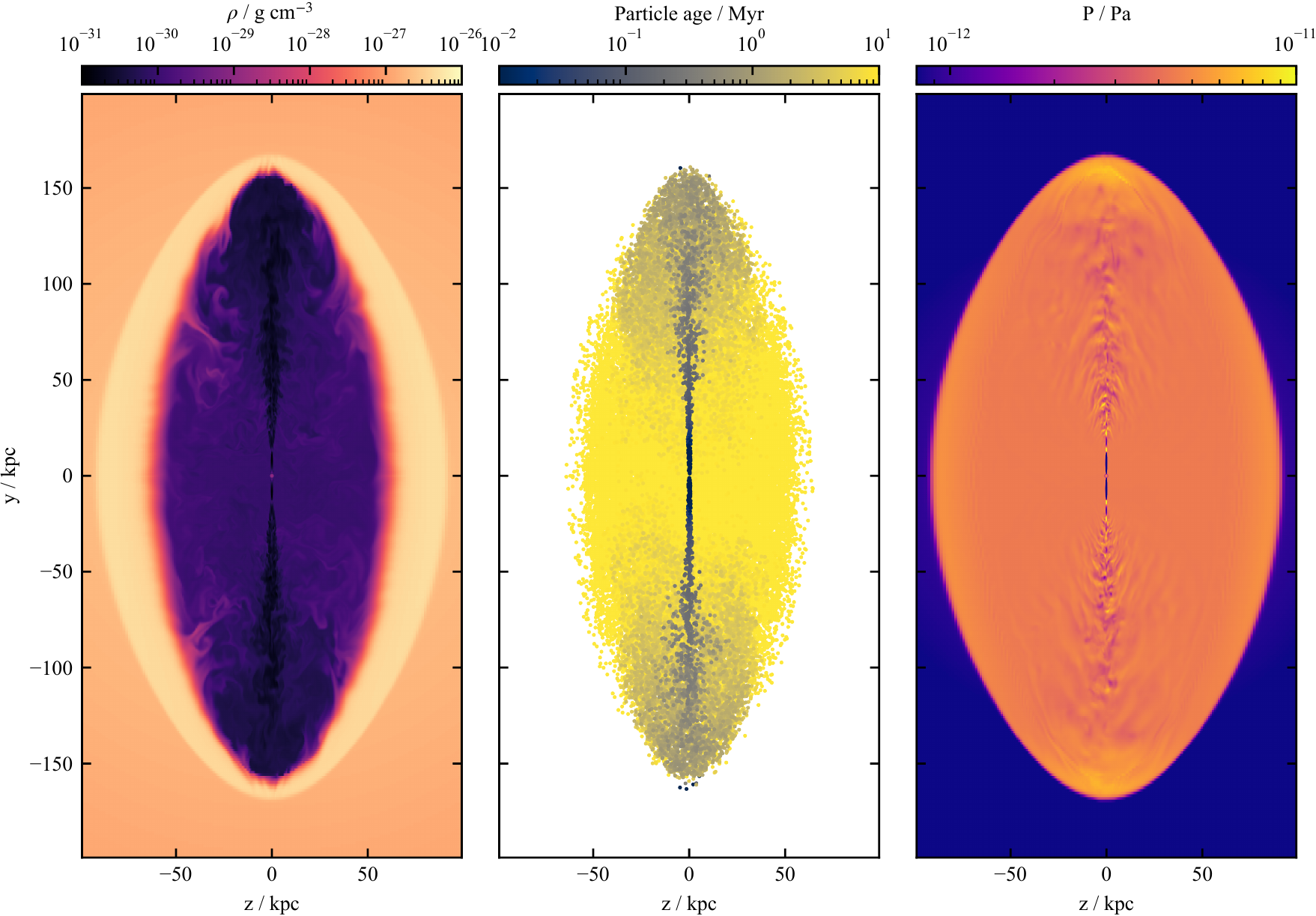}
    \caption{
      Mid-plane slices of fluid and particle quantities for the $0\rcore$-offset simulation at $t=32\myr$.
      \textit{Left}: Density.
      \textit{Centre}: Lagrangian tracer particles, coloured according to particle age since they were last shocked.
      \textit{Right}: Pressure.
    }
    \label{fig:sr_density-particle-distribution}
  \end{figure*}

  \begin{figure*}
    \centering
    \includegraphics{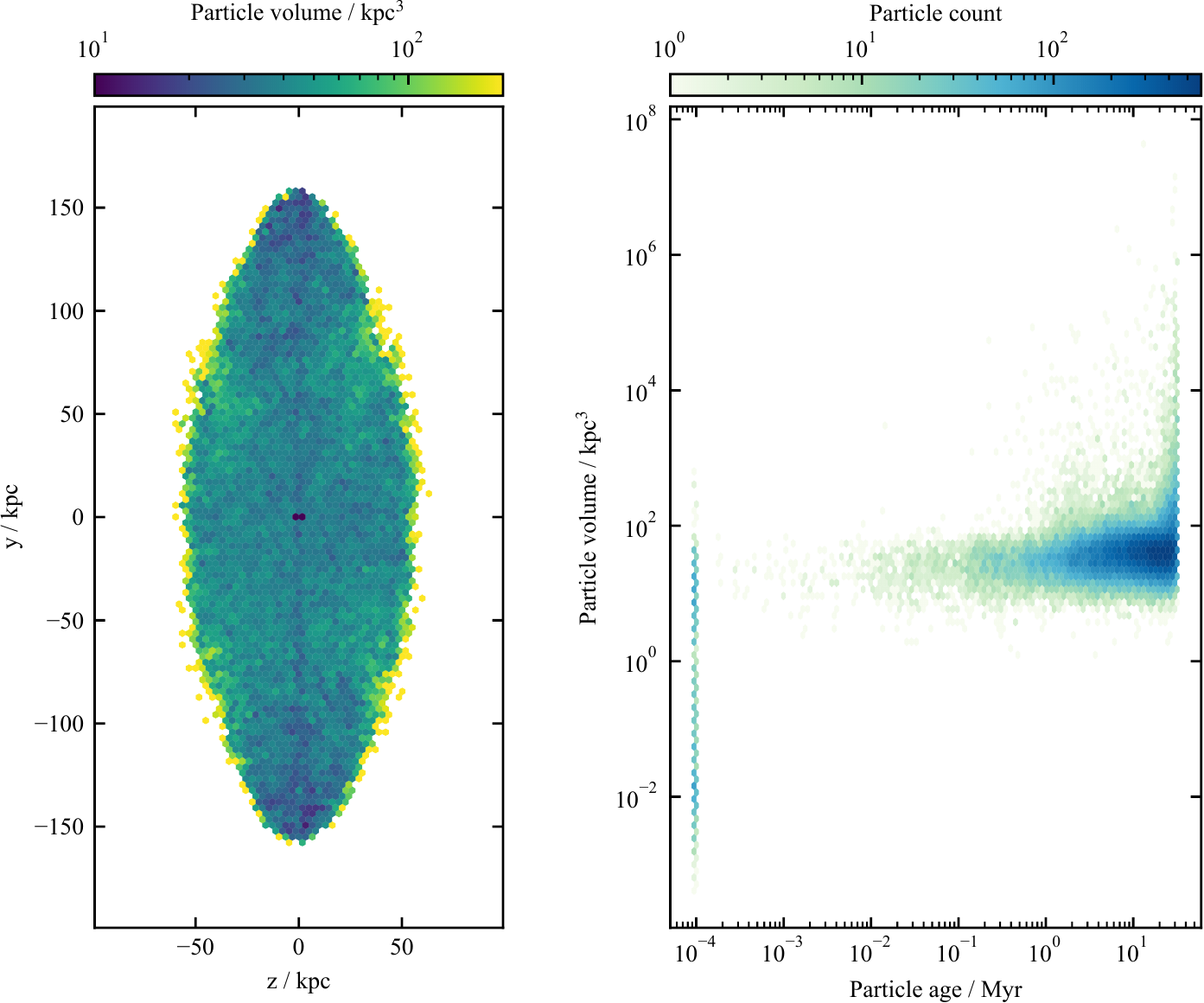}
    \caption{
      The volume occupied by the Lagrangian tracer particles for the $0\rcore$-offset simulation at $t=32\myr$.
      Particle volumes are calculated using a Voronoi tessellation; open regions at the edges of the jet cocoon are removed from the calculations.
      \textit{Left}: Hexagonally binned median particle volume as a function of position within the jet.
      \textit{Right}: Hexagonally binned particle volume as a function of age since last shock.
    }
    \label{fig:sr_particle_filling_factor}
  \end{figure*}

  In this section, we describe our novel implementation for calculating spatially resolved radiative losses in hydrodynamic simulations; we make use of version $4.3$ of the freely available numerical simulation code \textsc{pluto}\footnote{\url{http://plutocode.ph.unito.it/}} \citep{Mignone2007}.
  Our approach uses the newly introduced Lagrangian particle module in \textsc{pluto} $4.3$ to inject tracer particles that are advected with the fluid, and record particle acceleration in strong shocks and subsequent radiative and adiabatic losses.
  The particle pressure and shock histories are recorded during the simulations and used in post-processing to obtain lossy emissivities per particle, based on the \textit{Radio AGN in Semi-analytic Environments} (RAiSE) method presented by \citet[hereafter RAiSE~II]{Turner2018a}.
  These emissivities are converted into surface brightness maps by integrating the particle emissivities along the observer's line-of-sight, assuming that the radio source is optically thin; this is a reasonable assumption for kpc-scale jets and lobes, which dominate the emissivity in our simulations.

  In \cref{sec:sr_tracer_particles} we describe the role of the Lagrangian tracer particles and their technical implementation.
  We describe the post-processing procedure in \cref{sec:sr_raise_equations}, and the production of surface brightness maps in \cref{sec:sr_surface_brightness}.

  \subsection{Lagrangian tracer particles}
  \label{sec:sr_tracer_particles}
  
    In the PRAiSE framework, tracer particles are injected into the fluid with sufficient temporal frequency to sample the region of interest, using the Lagrangian particle module in \textsc{pluto} $4.3$.
    Each particle records the local fluid properties using Triangular Shape Cloud interpolation weights: at each timestep, the tracer particles are assigned fluid quantities based on the weighted grid quantities of the three closest cells in each dimension.
    Shocked zones are detected on the simulation grid using the flagging scheme described by \citet[][Appendix B]{Mignone2012}; in brief, a zone is flagged as shocked if the divergence of the velocity is negative, $\nabla \cdot \vec{v} < 0$, and the local pressure gradient exceeds a given threshold $\epsilon_p$.
    Multiple pressure thresholds can be specified for a given simulation, allowing different shock strengths to be included in post-processing.
    For each particle we record the last time it was in a shocked zone, for the given shock threshold.
    The fluid is evolved on an Eulerian grid, and the particles are advected according to the local fluid velocity, using the same time-marching scheme as the fluid quantities.

    In \cref{fig:sr_shock_threshold} we show the effect of different pressure thresholds on shocks captured in one of the jet simulations we present in \cref{sec:sr_validation}.
    Each group of two panels shows a plot of the lobe and jet tracer particles (based on a velocity cut of $0.3c$, discussed further in \cref{sec:sr_spatially_resolved_losses}), coloured by their age, for different pressure thresholds, corresponding to Mach number thresholds of $\mathcal{M} \sim 1.02, 1.18, 2.24$.

    We find that the lowest threshold captures shocked particles not just along the jet beam and at the terminal shock, but throughout the backflow as well.
    For the middle threshold, weaker shocks are captured at the very edges of the lobes, but not the shocks in the backflow.
    The highest threshold captures only strong shocks within the collimated jet, and at the jet head for the jet at early times, when the lobe expansion is fast.

    For the rest of this work, we consider only a single threshold, $\epsilon_p = 5$, corresponding to $\mathcal{M}\sim2.24$.
    While we can detect weaker shocks, particle acceleration theory predicts that our chosen Mach number is the critical Mach number below which particle acceleration is unlikely to occur in many situations \citep{Vink2014,Kang2019,Ha2022}.
    A similar threshold of $\epsilon_p \sim 3$ is chosen by \citet{Vaidya2018} for the same purpose of detecting shocks for particle acceleration.
    The electron energy evolution approach outlined below in \cref{sec:sr_raise_equations} allows for an electron injection index that varies with shock strength, however, we have chosen a simple single injection index and injection only at strong shocks for this initial application.

    The interpolated particle density and pressure are smoothed in time to reduce sampling noise caused by the interpolation process.
    A Savitzky--Golay filter is used, with polynomial order 3 and window length 5 \citep{Savitzky1964}.
    This filter preserves the overall trend in hydrodynamic quantities, which is most important for calculating losses over the particle lifetime.
    The particle properties are sampled every $0.01\myr$, corresponding to a smoothing time-span of $0.05\myr$.
    In \cref{fig:sr_particle_smoothing} the smoothed and unsmoothed pressure histories for several particles are shown in the left panel.
    Meanwhile, the right panel shows the effect of two different smoothing window lengths on the emissivity evolution of the same particles at an observing frequency of $9.0\,\textrm{GHz}$ and redshift $z=0.05$, calculated using the equations described below in \cref{sec:sr_raise_equations}.
    A high observing frequency is chosen to highlight the prominence of radiative losses.

    We find that our chosen smoothing parameters capture the overall pressure trends, and retain local features.
    A window length of $5$ accurately captures the full particle emissivity evolution, while removing sharp pressure discontinuities.

    Each particle is taken to represent a distinct packet of electrons.
    This approach is valid assuming that the electrons obey the same transport equations as the fluid.
    While this assumption is not correct in the presence of strong shocks, it holds in the presence of smoother flows \citep{Tregillis2001,Mimica2009}.
    We restrict our focus solely to the post-shock evolution of an electron population, and so this assumption is sufficient to describe their dynamics.

  \subsection{Synchrotron emissivity}
  \label{sec:sr_raise_equations}

    The post-processing approach is based on the work presented by \citetalias{Turner2018a} for calculating the evolution of electron energy losses using an analytic iterative approach. 
    Here we first summarise the theoretical approach of RAiSE, and present the modifications necessary to apply the model to numerical simulations; the reader is referred to \citetalias{Turner2018a} for the full derivation.

    We use the standard approximation \citep{Kaiser1997a,Longair2011} that the electrons radiate the bulk of their energy at a critical frequency $\nu_\textrm{c}$, related to the Larmor frequency as $\nu_\textrm{c} \approx \gamma^2 \nu_\textrm{L}$.
    Given $\nu_\textrm{L} = 3eB/2\pi m_\textrm{e}$, the Lorentz factor of electrons emitting at a frequency $\nu$ is given by the electron mass $m_\textrm{e}$, charge $e$ and magnetic field strength $B$ as
    \begin{equation}
      \gamma(\nu) = \sqrt{\frac{ 2 \pi \nu m_\textrm{e} }{ 3 e B }}\,.
      \label{eqn:sr_larmor_lorentz_factor}
    \end{equation}

    As the electrons move through the fluid, their energy distribution evolves due to loss processes and re-acceleration at shocks.
    Assuming that the electron population has a power law energy distribution $N(E, t)dE = N_0 E^{-s} dE$ after acceleration at a strong shock, the effect of losses on the electron population can be modelled.
    Adiabatic losses (for a packet of electrons expanding adiabatically as $dV_\textrm{packet} \propto t^{-a_p / \Gamma_\textrm{c}}$ for a cocoon adiabatic index $\Gamma_\textrm{c}$), synchrotron radiative losses, and losses due to the up-scattering of cosmic microwave background (CMB) photons are included in the following equations.
    The parameter $a_p$ relates to how the cocoon pressure evolves with time, $p \propto t^{a_p}$, and is calculated iteratively at each timestep $n$ as $a_p(t_{n-1}, t_{n}) = \log(p_n/p_{n-1}) / \log(t_n/t_{n-1})$.
    Following \citetalias{Turner2018a}, we calculate the Lorentz factor at the time each particle (packet of electrons) was last accelerated, $t_\textrm{acc}$.
    This Lorentz factor, $\gamma_\textrm{acc}$, is calculated from the emitting Lorentz factor $\gamma(\nu)$ at the current time using an iterative method, where the Lorentz factor evolves as
    \begin{equation}
      \gamma_n = \frac{ \gamma_{n - 1} t_n^{a_p(t_{n-1}, t_n) / 3 \Gamma_\textrm{c}} }{ t_{n - 1}^{a_p (t_{n-1}, t_n)/ 3\Gamma_\textrm{c}} - a_2(t_{n-1}, t_n) \gamma_{n - 1} }\,,
      \label{eqn:sr_praise_iterative_lorentz}
    \end{equation}
    for time $t$ decreasing from $(t_0,\gamma_0)=(t,\gamma)$ through to $(t_N, \gamma_N)=(t_\textrm{acc}, \gamma_\textrm{acc})$, the moment the electron population was accelerated.
    If a population of electrons experiences heavy losses it will lack the energy to radiate above a cut-off frequency; when this occurs $\gamma_\textrm{acc}$ will rapidly approach infinity, indicating a lack of emission for the chosen frequency; the emissivity for this electron population is set to zero.
    The parameter $a_2(t_{n-1}, t_n)$ depends on the local magnetic field energy density $u_\textrm{B}$ and the CMB energy density $u_\textrm{C}$,
    \begin{multline}
      a_2(t_{n-1}, t_n) =\\
      \frac{ 4 \sigma_\textrm{T} }{ 3 m_\textrm{e} c } \left[ \frac{ u_\textrm{B} (t_n) }{ a_3 } t_n^{-a_p} \left( t_{n-1}^{a_3} - t_n^{a_3} \right) + \frac{ u_\textrm{C} }{ a_4 } \left( t_{n-1}^{a_4} - t_n^{a_4} \right) \right]\,,
      \label{eqn:sr_praise_a2}
    \end{multline}
    with Thomson cross section $\sigma_\textrm{T}$ and speed of light $c$.
    The parameters $a_3 = 1 + a_p ( 1 + 1/3\Gamma_\textrm{c} )$ and $a_4 = 1 + a_p / 3\Gamma_\textrm{c}$ both depend on how the local lobe pressure changes with time.
    Energy losses increase rapidly with redshift $z$ due to an increasing CMB energy density, as $u_C = 4.00\times 10^{-14} (1 + z)^4\,\textrm{J m}^{-3}$.

    Magnetic fields are not included self-consistently in the simulations presented in this work.
    We made this initial simplification primarily for comparison with previous analytic work.
    However, self-consistently simulated magnetic fields can have important dynamical effects.
    A random or helical magnetic field configuration in the jet leads to a configuration predominantly aligned with the jet, making the radio emission anisotropic \citep{Huarte-Espinosa2011,Hardcastle2014}.
    Magnetic fields are intermittent and can vary considerably for a given lobe pressure \hbox{\citep[e.g.,][]{Gaibler2009}}.
    While they may stabilise contact surfaces around radio lobes against instabilities \citep{Gaibler2009}, they may also make jets more unstable \citep{Mignone2010,Mukherjee2020}.
    Particle transport is also known to be affected by magnetic fields, with strong suppression only perpendicular to the field lines \citep[e.g.,][]{Owen2022}.
    Our assumption that particles only move with the bulk flow effectively corresponds to a tangled-magnetic-field approach.
    Studies of the mixing of the populations of relativistic electrons in radio lobes seem to suggest this is a reasonable approximation \citep[e.g.,][]{Turner2018a}.
    Bearing the above caveats in mind, our simulations should be useful to explore general radio source properties.

    Therefore, a mapping between magnetic field energy density and a hydrodynamical quantity is required.
    Following \citet{Kaiser1997a}, the lobe pressure $p = (\Gamma_\textrm{c} - 1)(u_\textrm{e} + u_\textrm{B} + u_\textrm{T})$ is a function of the electron, magnetic field, and thermal energy densities.
    Using the ratio of electron to magnetic field energy density, $\eta = u_\textrm{B}/u_\textrm{e}$ (referred to herein as the equipartition factor), and assuming that thermal particles contain negligible energy ($u_\textrm{T} \sim 0$), the magnetic energy density and hence magnetic field strength can be written in terms of pressure as
    \begin{equation}
      B = \left( \frac{ 2 \mu_0 p }{ \Gamma_\textrm{c} - 1} \frac{ \eta }{ \eta + 1 } \right)^{1/2}\,.
    \end{equation}

    The particle rest-frame emissivity per unit volume and per unit solid angle for a specific frequency $\nu '$ is given by 
    \begin{multline}
      j_{\nu'}' = \frac{K(s)}{4 \pi} (\nu')^{(1-s)/2} \frac{ \eta^{(s+1)/4} }{ (\eta + 1)^{(s + 5)/4} }\\
      \times p(t)^{(s+5)/4} \left[ \frac{ p(t_\textrm{acc}) }{ p(t) } \right]^{ 1 - 4/(3 \Gamma_\textrm{c}) } \left[ \frac{ \gamma_\textrm{acc} }{ \gamma } \right]^{2 - s}\,,
      \label{eqn:sr_praise_emissivity}
    \end{multline}
    for equipartition factor $\eta$, electron energy power law exponent $s$, and cocoon adiabatic index $\Gamma_\textrm{c}$.
    Here, $p(t)$ and $p(t_\textrm{acc})$ are the local particle pressures at the current time and time of acceleration respectively, while $\gamma$ and $\gamma_\textrm{acc}$ are the corresponding Lorentz factors.
    $K(s)$ is the radio source specific constant,
    \begin{multline}
      K(s) = \frac{ \kappa(s) }{ m_\textrm{e}^{ (s+3)/2} c (s+1) } \left[ \frac{ e^2 \mu_0 }{ 2 (\Gamma_\textrm{c} - 1) } \right]^{(s + 5)/4}\\
      \times \left[ \frac{3}{\pi} \right]^{s/2} \left[ \frac{ \gamma_\textrm{min}^{2-s} - \gamma_\textrm{max}^{2-s} }{s-2} - \frac{ \gamma_\textrm{min}^{1-s} - \gamma_\textrm{max}^{1-s} }{1-2} \right]^{-1}\,,
      \label{eqn:sr_radio_source_constant}
    \end{multline}
    which depends on the vacuum permeability $\mu_0$, and accelerated electron Lorentz limits $\gamma_\textrm{min}$, $\gamma_\textrm{max}$.
    The constant $\kappa (s)$ is given \citep{Longair2011} as
    \begin{equation}
      \kappa (s) = \frac{ \Gamma \left( \frac{s}{4} + \frac{19}{12} \right) \Gamma \left( \frac{s}{4} - \frac{1}{12} \right) \Gamma \left( \frac{s}{4} + \frac{5}{4} \right)}{ \Gamma \left( \frac{s}{4} + \frac{7}{4} \right) }\,.
      \label{eqn:sr_synchrotron_kappa}
    \end{equation}

    The transformation from the fluid rest frame (primed quantities) to the observer frame (unprimed quantities) is achieved by
    \begin{equation}
      j_\nu = D^{2 + \alpha} j_{\nu'}',
      \label{eqn:sr_praise_observer_emissivity}
    \end{equation}
    where $\alpha$ is the spectral index, and $D=1/(\gamma [1 - \vec{\beta} \cdot \vec{n}])$ is the relativistic Doppler factor given by the bulk Lorentz factor $\gamma$, the bulk 3-velocity of the fluid $\vec{\beta}$, and the observing normal $\vec{n}$.
    The spectral index at injection time is used for this transformation, assuming no losses; in practice, we find that particles with high bulk Lorentz factors are in general recently shocked, and so losses are negligible.
    For radio sources with a redshift $z > 0$, the observing frequency is related to the emitting frequency as $\nu_{e} = (1 + z)\nu_\textrm{0}$.
    We apply this redshift dependence as well as the Lorentz transformation due to the bulk velocity to the observing frequency to obtain the emitting frequency in the plasma rest-frame, which is then used in Eq.~\ref{eqn:sr_praise_emissivity}; all frequencies stated in this work are observed frequencies.
    
  \subsection{Ray-traced surface brightness}
  \label{sec:sr_surface_brightness}

    Following the procedure described in \cref{sec:sr_raise_equations}, the emissivity per unit volume for each Lagrangian tracer particle is calculated for a specific point in time, using the particle history to account for radiative and adiabatic loss processes.
    This emissivity is integrated over a two-dimensional observing grid to produce a surface brightness map.

    The surface brightness for each pixel in the observing grid is calculated by casting rays through the entire simulation volume, perpendicular to the observing grid.
    Each ray is partitioned into discrete elements of length $\Delta_r$ along the line-of-sight.
    A k-d tree, an effective data structure for binary spatial partitioning of a dataset, is constructed from particle positions to enable fast nearest-neighbour lookups for a given coordinate; with it, the closest Lagrangian tracer particle to each ray element is found.
    The volume emissivity of this particle is then assigned to the corresponding ray element.
    The total surface brightness for a given pixel is then calculated as $B_\nu = \int j_\nu (r) \textrm{d}r$ using a line-of-sight integral along the ray.

    In this work we choose $\Delta_r = 0.1\kpc$; this is sufficient to sample the particles, as discussed in \cref{sec:sr_dynamics}.
    The surface brightness maps are convolved with a two-dimensional Gaussian beam with $5\kpc$ full width at half-maximum (FWHM), corresponding to a $5\,\textrm{arcsec}$ beam at redshift $z=0.05$.

\section{Validation}
\label{sec:sr_validation}

  \begin{figure*}
    \centering
    \includegraphics{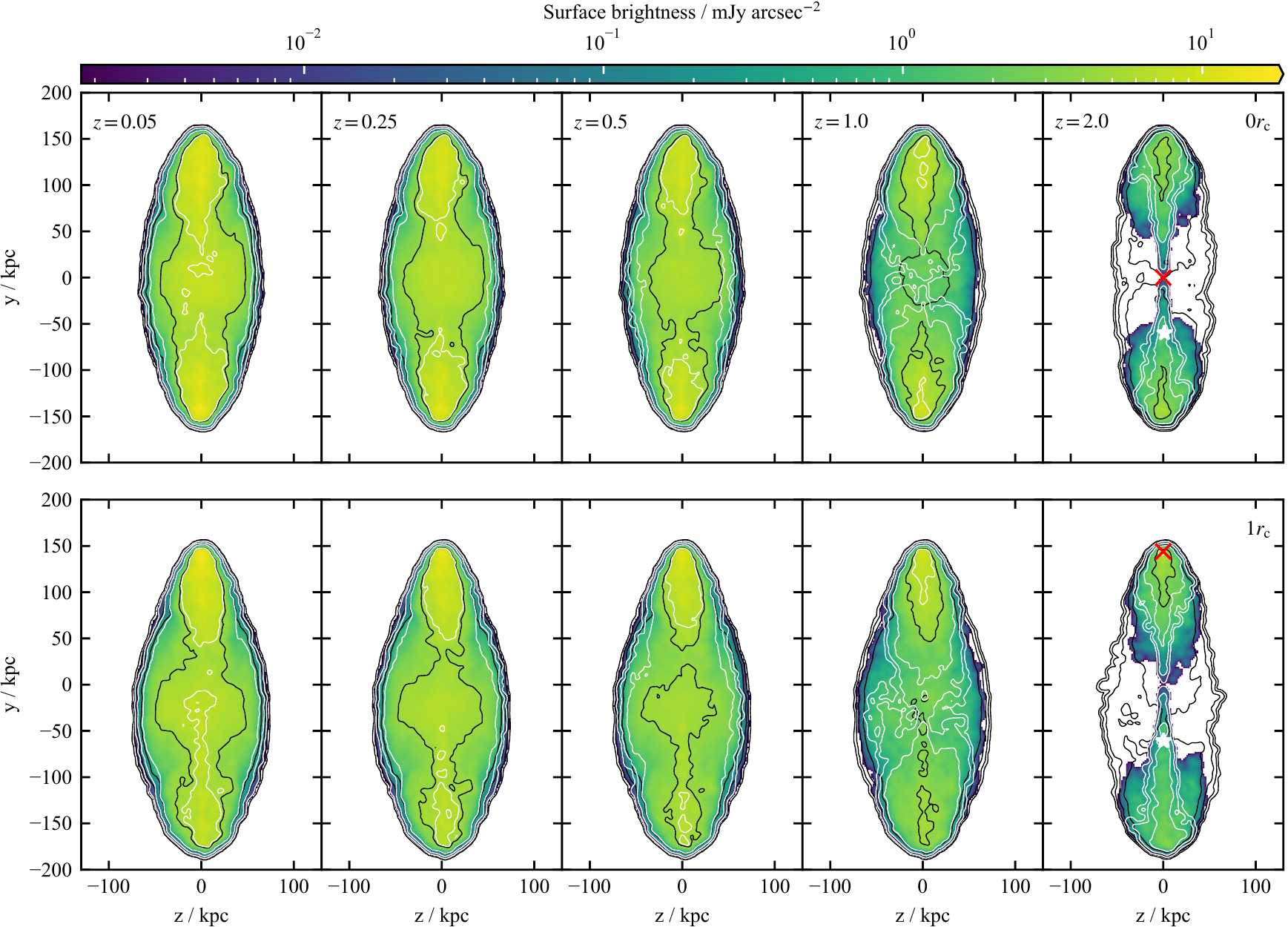}
    \caption{
      Synthetic surface brightness maps at a range of redshifts at $32\myr$.
      Identical physical pixel ($1.8\kpc$) and beam ($5\kpc$) sizes are used for all redshifts.
      Surface brightness maps for the $0\rcore$-offset simulation are plotted in the top row, while the $1\rcore$-offset simulation maps are plotted in the bottom row.
      The location of the environment centre for each simulation is marked with the red cross in the final panel of each row.
      The white stars mark the location where local spectra within the jet is plotted in \cref{fig:sr_local_spectra}.
      The colour map shows $1.4\,\textrm{GHz}$ surface brightness in units of mJy arcsec$^{-2}$, with limits corresponding to $[5\times 10^{-2}, 5\times 10^{2}]\,\textrm{mJy beam}^{-1}$ at $z=0.05$.
      The black contours are plotted for the $150\,\textrm{MHz}$ emission, while the white contours are plotted for the $9.0\,\textrm{GHz}$ emission.
      All plots have the same contour levels in $\textrm{mJy arcsec}^{-2}$, five logarithmically spaced between the lower and upper surface brightness colour map limits; at $z=0.05$ these correspond to $0.05, 0.5, 5, 50, 500\,\textrm{mJy beam}^{-1}$.
    }
    \label{fig:sr_radio-morphology}
  \end{figure*}

  \begin{figure}
    \centering
    \includegraphics{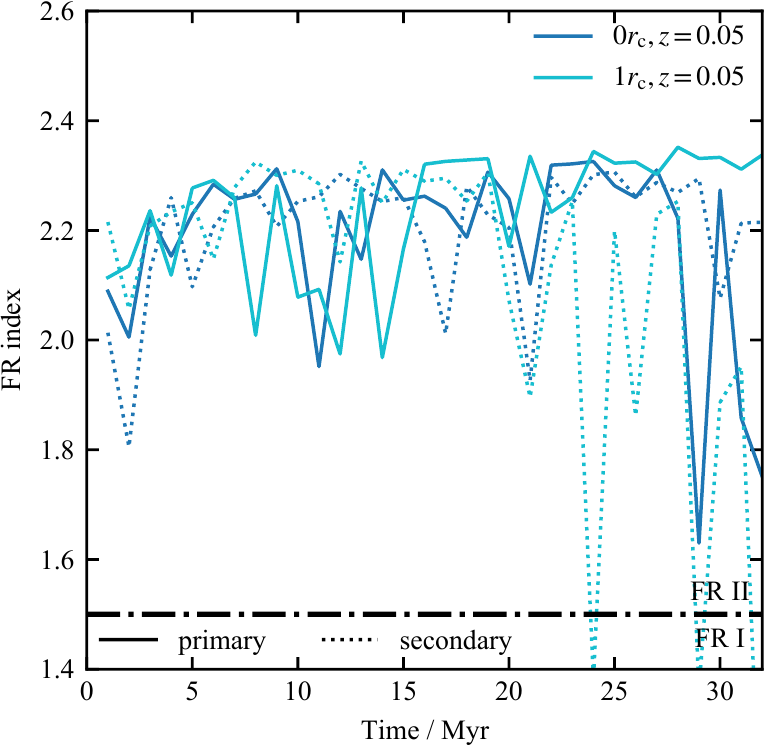}
    \caption{
      The FR index at $z=0.05$.
      Primary lobes are plotted as solid lines, secondary lobes as dotted lines.
      The theoretical transition point between FR~I and FR~II morphology as defined by the FR index is plotted at $y=1.5$ as the black dot-dashed line.
    }
    \label{fig:sr_fr_index}
  \end{figure}

  \begin{figure*}
    \centering
    \includegraphics{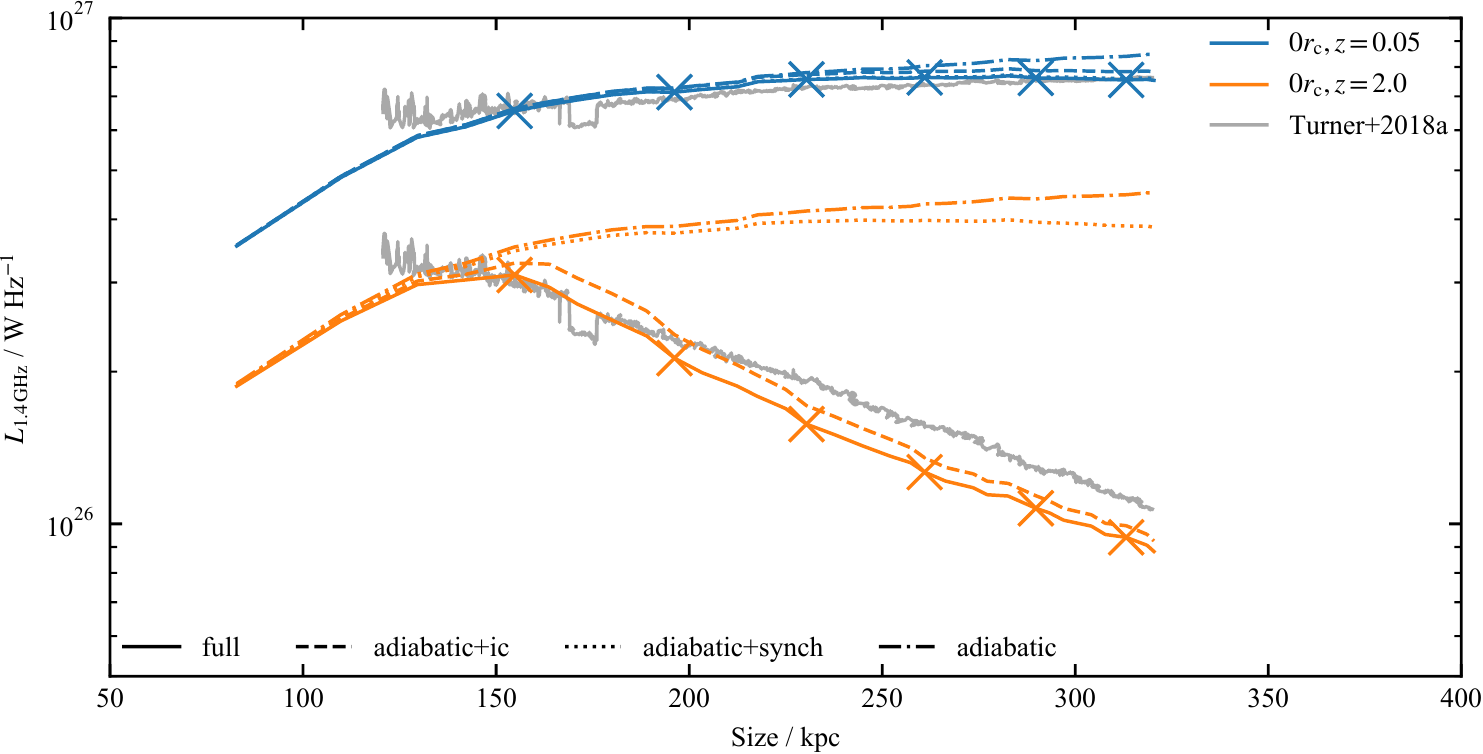}
    \caption{
        Tracks of the $0\rcore$-offset simulation through the size-luminosity diagram.
        The total luminosity as a function of different loss processes is plotted with different line styles: the solid lines have all losses enabled, the dashed and dotted lines model adiabatic and either inverse-Compton or synchrotron losses respectively, while the dash-dot line only has adiabatic losses enabled.
        The crosses mark $5\myr$ increments for both tracks.
        A theoretical luminosity calculation following the method presented in \citetalias{Turner2018a} is plotted for comparison as the grey curve at both $z=0.05$ (upper) and $z=2.0$ (lower).
        The main effect that reduces the luminosity at the higher redshift is the energy loss due to inverse-Compton scattering.
    }
    \label{fig:sr_pd_tracks_losses}
  \end{figure*}

  \begin{figure*}
    \centering
    \includegraphics{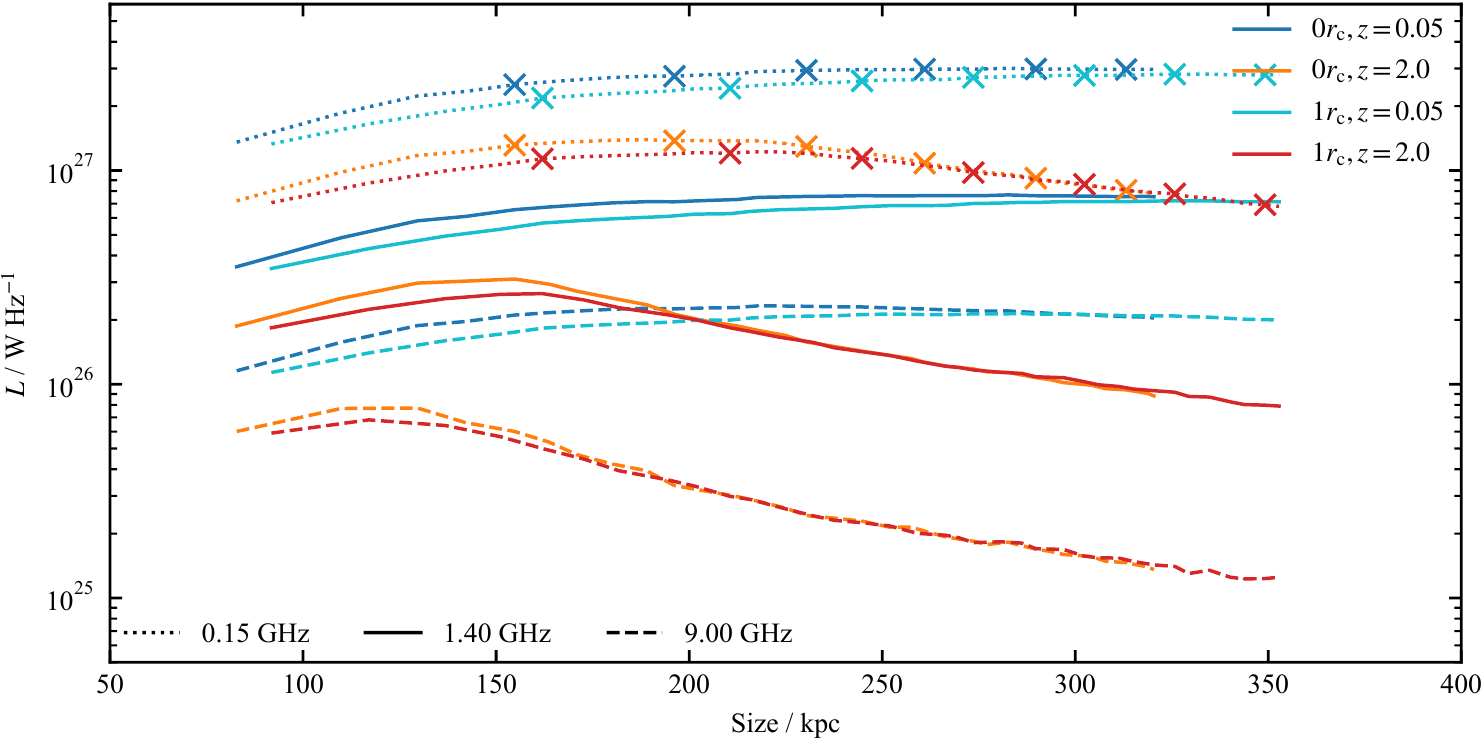}
    \caption{
        Frequency and redshift dependence of size-luminosity tracks.
        Total luminosities for the different frequencies plotted in \cref{fig:sr_radio-morphology} are plotted with different line styles: dotted lines are for $\nu = 150\,\textrm{MHz}$, solid lines for $\nu = 1.4\,\textrm{GHz}$, and dashed lines for $\nu=9.0\,\textrm{GHz}$.
        The crosses mark $5\myr$ increments.
    }
    \label{fig:sr_pd_tracks_frequencies}
  \end{figure*}

  \subsection{Simulations}\label{sec:sr_tech-setup}

    The simulations used in this work were first presented in \citet[hereafter Paper~I]{Yates-Jones2021}; a full description of the technical details is presented therein.
    Briefly, \textsc{pluto} version $4.3$ was used with the relativistic hydrodynamics module to solve the fluid conservation equations on a three-dimensional Cartesian grid with $2^\textrm{nd}$-order Runge-Kutta time-stepping, the HLLC Riemann solver, linear reconstruction, and the \textsc{minmod} limiter in the presence of shocks.
    The Taub-Mathews \citep{Mathews1971,Mignone2007a} equation of state is used to relate fluid quantities, and the Lagrangian particle module is used to inject tracer particles with the jet.

    We simulate radio sources in both spherically symmetric and asymmetric environments.
    The initial gas density distribution is set according to the radial isothermal beta profile \citep{King1962,Cavaliere1976}, modified to offset the jet injection region from the centre of the cluster:
    \begin{equation}
      \rho(r) = \rho_0 \left[ 1 + \left( \frac{r}{r_\textrm{c}} \right)^2 \right] ^ {-3 \beta / 2}\,.
    \end{equation}
    The pressure is given as $P(r) = \frac{k_\textrm{B} T \rho(r)}{\mu m_\textrm{H}}$.
    For a dark matter halo mass of $M_\textrm{halo} = 3 \times 10^{14}\,\textrm{M}_{\sun}$ and $\beta = 0.38$, typical of clusters, the core radius is $\rcore = 144\kpc$, given $\rcore = 0.1 R_\textrm{vir}$ as in \citet{Hardcastle2013}.
    The necessary gravitational acceleration to ensure a stable environment is derived under the assumption of hydrostatic equilibrium.

    Two simulations are considered in this paper.
    Both have a one-sided total relativistic power of $Q=3\times10^{38}\,\textrm{W}$, and are offset either $0$ or $1$ core radii (i.e. $144\kpc$) from the cluster centre.
    Throughout this paper the simulations are referred to as $0\rcore$-offset and $1\rcore$-offset respectively; they correspond to the ``off0r'' and ``off1r'' simulations in \citetalias{Yates-Jones2021}.
    As in \citetalias{Yates-Jones2021}, the jet propagating towards (away from) the cluster centre is referred to as the primary (secondary) jet.
    Initially, the jet has a half-opening angle of $\theta_\textrm{j} = 10.0^\circ$ and is relativistic, $\gamma_\textrm{j} = 5$, where $\gamma_\textrm{j} = 1 / \sqrt{1 - v_\textrm{j}^2/c^2}$ is the jet bulk Lorentz factor.
    The jet injection region is defined as in \citetalias{Yates-Jones2021}: a sphere centred at the origin, with radius $r_0 = 1\kpc$.
    The fluid pressure and density within the injection zone are continuously overwritten based on the desired jet values ($P_\textrm{j}$, $\rho_\textrm{j}$) as
    \begin{align}
      \rho_\textrm{i}(r) &= 2 \rho_\textrm{j} (1 + (r / r_0)^2)^{-1}\\
      P_\textrm{i}(r) &= 2^\Gamma P_\textrm{j} \left(\frac{\rho(r)}{2\rho(r_0)}\right)^\Gamma
      \,,
    \end{align}
    for an ideal adiabatic index $\Gamma=5/3$, valid for the kinetically dominated jet material at the inlet.
    The velocity is set to the jet velocity $v_\textrm{j}$ within a cone defined by $|\theta| \le \theta_\textrm{j}$, and $0$ elsewhere.

    Lagrangian tracer particles are uniformly scattered throughout the jet injection cone to ensure that the jet is well sampled; two particles (one per jet side) are injected every $\sim 1\,\textrm{kyr}$.
    These particles are advected along the fluid streamlines and sample the radio source cocoon.
    The particles function purely as tracers; particle back-reaction on the fluid is not modelled in these simulations.
    In subsequent analysis, we use a single pressure threshold of $\epsilon_p = 5$ to track shocks, corresponding to a minimum Mach number of $\mathcal{M} \sim 2.24$.

  \subsection{Dynamics}
  \label{sec:sr_dynamics}

    In \cref{fig:sr_density-particle-distribution} we show the grid density and pressure for the $0\rcore$-offset simulation at the final simulation time, $t=32\myr$.
    On the same scale, we plot the spatial distribution of the Lagrangian tracer particles coloured by their age since they were last shocked.
    Particles are drawn in the order they were injected, so the most recently injected particles are drawn on top.
    These are found along the jet and towards the tip of the lobe.
    We note that the number of recently shocked particles at the very edges of the lobe tips is low, due to the turbulent disruption of the jet.
    There are, however, still sufficient numbers of recently shocked particles in the lobe head region to reproduce the observed surface brightness enhancement typical of FR~II radio sources; we discuss this further in \cref{sec:sr_validation_sb}.

    It is important to consider how well the tracer particles sample the jet cocoon.
    An under-sampled cocoon will produce unphysical features in the radio maps.
    We check that the cocoon is well-sampled by calculating the volume represented by each particle.
    First, the Voronoi tessellation of all particles is calculated.
    The Voronoi tessellation for a given set of seed points produces a set of regions enclosing the space closest to their seed point.
    Next, the volume of the convex hull corresponding to each Voronoi region is calculated.
    Particles at the edge of the cocoon formally have infinite volume, as their Voronoi region is open; however, to demonstrate how well the interior of the cocoon is sampled, those particles are excluded from the following volume calculations.
    \Cref{fig:sr_particle_filling_factor} demonstrates that our choice for the temporal frequency of particle injection is sufficient to adequately sample the cocoon.
    The left-hand panel in \cref{fig:sr_particle_filling_factor} shows the spatial distribution of particle volumes throughout the cocoon: the median volume is plotted for the $0\rcore$-offset simulation at $t=32\myr$, binned onto a two-dimensional hexagonal grid.
    The edge of the cocoon has high median volumes, however, the cocoon interior is largely sampled by particles with volumes less than $100\kpc^3$, comparable to a resolution of $5\kpc$.
    At the lobe tips and along the jets, the median particle volume is lower than in the rest of the cocoon, as expected.
    This demonstrates that the method results in a reasonably uniform sampling with a spatial resolution of about $5\kpc$, justifying the grid resolution of $1.8\kpc$ in the synthetic images below.

    The hexagonally binned scatter plot shown in the right-hand panel of \cref{fig:sr_particle_filling_factor} shows the number of particles with a given volume as a function of their age.
    Two features stand out in this plot.
    First, the most recently shocked particles have significantly smaller volumes than the rest of the particle population; these are particles travelling along the jet, in close proximity with each other.
    Second, the majority of all particles have a volume of less than $100\kpc^3$ (effective resolution of $\sim 5\,\kpc$), regardless of age.
    Only a small population of the oldest particles (least recently shocked, and hence least likely to contribute significantly to the integrated emissivity) have an effective resolution worse than that.

  \subsection{Surface brightness}
  \label{sec:sr_validation_sb}

    Following the process outlined in \cref{sec:sr_raise_equations}, we calculate the emissivity corresponding to each Lagrangian tracer particle at five redshifts (spanning $z=0.05 - 2.0$) and eleven observing frequencies (from $150\,\textrm{MHz}$ to $50\,\textrm{GHz}$).
    The radio source is assumed to be oriented in the plane of the sky for all surface brightness maps.

    An injection index of $s=2.2$ is used to initialise the electron energy population whenever a particle is shocked; this is consistent with the range of observed FR~II spectra in both hotspots and lobes \citep{Mahony2016,Harwood2017}.
    In this work we use a constant injection index and select only strong shocks with $\epsilon_p=5$; however an injection index coupled to shock Mach number is supported by our approach.
    The minimum Lorentz factor is chosen to be $\gamma_\textrm{min}=500$, consistent with values of several hundred found in observations of FR~II hotspots \citep{Hardcastle1998a,Godfrey2009,Turner2019}, and the maximum Lorentz factor is $\gamma_\textrm{max}=10^5$; these values are the same for all shocks.
    We choose the ratio between actual and equipartition magnetic field strengths to be $B/B_\textrm{eq}=0.4$, consistent with estimates from dynamical models \citep{Turner2018} and observational studies \citep{Ineson2017} of FR~II radio galaxies.
    This gives an electron to magnetic field energy density ratio of $\eta = (B_\textrm{obs}/B_\textrm{eq})^{(s+5)/2} = 0.03$ \citep{Croston2005}.
    We note that re-acceleration at weak shocks might well produce complex spectra.
    In this work, we only trace stronger shocks where we assume the electron energy distributions to be reset to a power law.

    In \cref{fig:sr_radio-morphology} we show synthetic surface brightness images for both simulations at five redshifts (increasing left to right) and three different observing frequencies.
    The colour map shows the surface brightness in mJy arcsec$^{-2}$ at $1.4\,\textrm{GHz}$, while the black contours show the surface brightness at $150\,\textrm{MHz}$--similar to the LOFAR Two-metre Sky Survey \citep[LoTSS,][]{Shimwell2017}; and the white contours show the surface brightness at $9.0\,\textrm{GHz}$--similar to the upper observing frequency of the GAMA Legacy ATCA Southern Survey \citep[GLASS,][]{Seymour2020}.
    The choice of mJy arcsec$^{-2}$ for surface brightness is to highlight the change in observed morphology with redshift.
    The surface brightness limits are chosen to correspond to $[5\times10^{-2} \le \textrm{SB} \le 5\times10^{2}\,\textrm{mJy beam}^{-1}]$ at $z=0.05$.
    While the limits are chosen to highlight the data, we note that the lower limit of $50\,\mu\textrm{Jy beam}^{-1}$ is comparable to the sensitivity of both LoTSS \citep[$\sim 100\,\mu\textrm{Jy beam}^{-1}$,][]{Shimwell2017} at $150\,\textrm{MHz}$, and GLASS \citep[$\sim 40\,\mu\textrm{Jy beam}^{-1}$,][]{Seymour2020} at $9.5\,\textrm{GHz}$, for similar beam sizes at $z=0.05$.

    Clear FR~II morphology is observed in both simulations, including a bright region near the tip of the lobes.
    These bright regions that are reminiscent of hotspots observed in FR~II radio sources are significantly narrower than the full low surface brightness extent; this is due to electron ageing, rather than a dynamical effect.
    The radio lobes have increased surface brightness near the tips, with a decrease towards the jet base.
    This is also due to the modelled loss processes: electrons in the equatorial regions are the oldest, and hence have the weakest emission despite the comparatively larger total emitting volume.
    
    As the redshift increases, the observed source morphology changes due to increased inverse-Compton losses.
    Older populations of electrons away from the jet head are no longer emitting at $1.4$ and $9.0\,\textrm{GHz}$, but are visible in the $150\,\textrm{MHz}$ contours.
    Despite this, the bright region near the tip of the lobes remains visible in all frequencies, across all redshifts.

    In \cref{fig:sr_fr_index} we show the Fanaroff-Riley (FR) index as a function of time, for both simulations, calculated at $z=0.05$.
    The FR index is a useful metric for classifying observed radio source morphology, and is defined following \citet{Krause2012} as $\textrm{FR}=2 x_\textrm{bright} / x_\textrm{length} + 1/2$ for each radio lobe.
    Surface brightness at $150\,\textrm{MHz}$ is used for the calculation, which is close to the $178\,\textrm{MHz}$ of the original definition.
    $x_\textrm{bright}$ is the radius of the brightest pixel in the lobe, and $x_\textrm{length}$ is the lobe length.
    A radio source with FR~I morphology will have an index of $0.5 < \textrm{FR} < 1.5$, while one with FR~II morphology will have an index of $1.5 < \textrm{FR} < 2.5$.
    The radio lobe length is defined as the distance to the furthest pixel from the jet core with a surface brightness within two dex of the maximum surface brightness.

    At almost all times, the radio lobes have an FR index identifying them as FR~IIs.
    Additionally, we find that this conclusion is largely independent of redshift.
    This is in agreement with the FR~II morphology produced--the hotspots are the brightest sections of an FR~II radio source, and hence fade the slowest with redshift, while the radio lobes experience strong losses.
    The primary $0\rcore$-offset lobe declines in FR index around $t=30\,\myr$.
    This occurs due to knots along the jet (as is visible in \cref{fig:sr_radio-morphology}), which are a transient feature.
    Despite this, the $0\rcore$-offset lobe is visually identifiable as having FR~II morphology.

  \subsection{Size-luminosity tracks}

    In \cref{fig:sr_pd_tracks_losses} we investigate the effect of different loss processes on evolutionary tracks through the size-luminosity diagram (also known as PD tracks) for the $0\rcore$-offset simulation, at redshifts $z=0.05$ and $z=2.0$.
    We plot for comparison the luminosity calculation following the method presented in \citetalias{Turner2018a} as the grey curves; the contribution of each particle to the total luminosity is calculated, and a Voronoi tessellation is used to assign each particle a volume.
    Both approaches agree well for large source sizes, however the ray-tracing approach used in this work better handles small particle numbers, demonstrated by the lack of jitter for small source sizes.

    The adiabatic track through the diagram is solely a function of the radio source dynamics; therefore, apart from the overall luminosity reduction due to the K-correction for the adopted spectral shape, no evolution with redshift is found.
    With only adiabatic losses, the track never turns over and luminosity continues increasing for larger source sizes.
    At low redshifts, the radiative synchrotron losses dominate, causing a luminosity turn-over in the full losses track at large sizes.
    However, at $z=2.0$ the inverse-Compton losses dominate the emissivity and cause a very sharp decline in total luminosity for large source sizes.
    These results are in agreement with previous analytical studies \citep[e.g.][]{Kaiser1997,Willott1999,Turner2015,Hardcastle2018} and demonstrate that PRAiSE is capturing the relative importance of different loss processes for a given redshift well.

    In \cref{fig:sr_pd_tracks_frequencies} the $z=0.05$ and $z=2.0$ evolutionary tracks through the size-luminosity diagram are plotted for both simulations at three different frequencies ($150\,\textrm{MHz}$, $1.4\,\textrm{GHz}$, and $9.0\,\textrm{GHz}$), with all loss mechanisms enabled.
    These three frequencies are the same frequencies for which the surface brightness is shown in \cref{fig:sr_radio-morphology}.
    Both synchrotron and inverse-Compton radiative losses increase as the observing frequency increases.
    As shown in \cref{fig:sr_pd_tracks_losses}, inverse-Compton losses dominate at $z=2.0$.
    This is reflected in the declining luminosity with size for all frequencies at high redshifts.

\section{Results}
\label{sec:sr_results}

  \begin{figure*}
    \centering
    \includegraphics{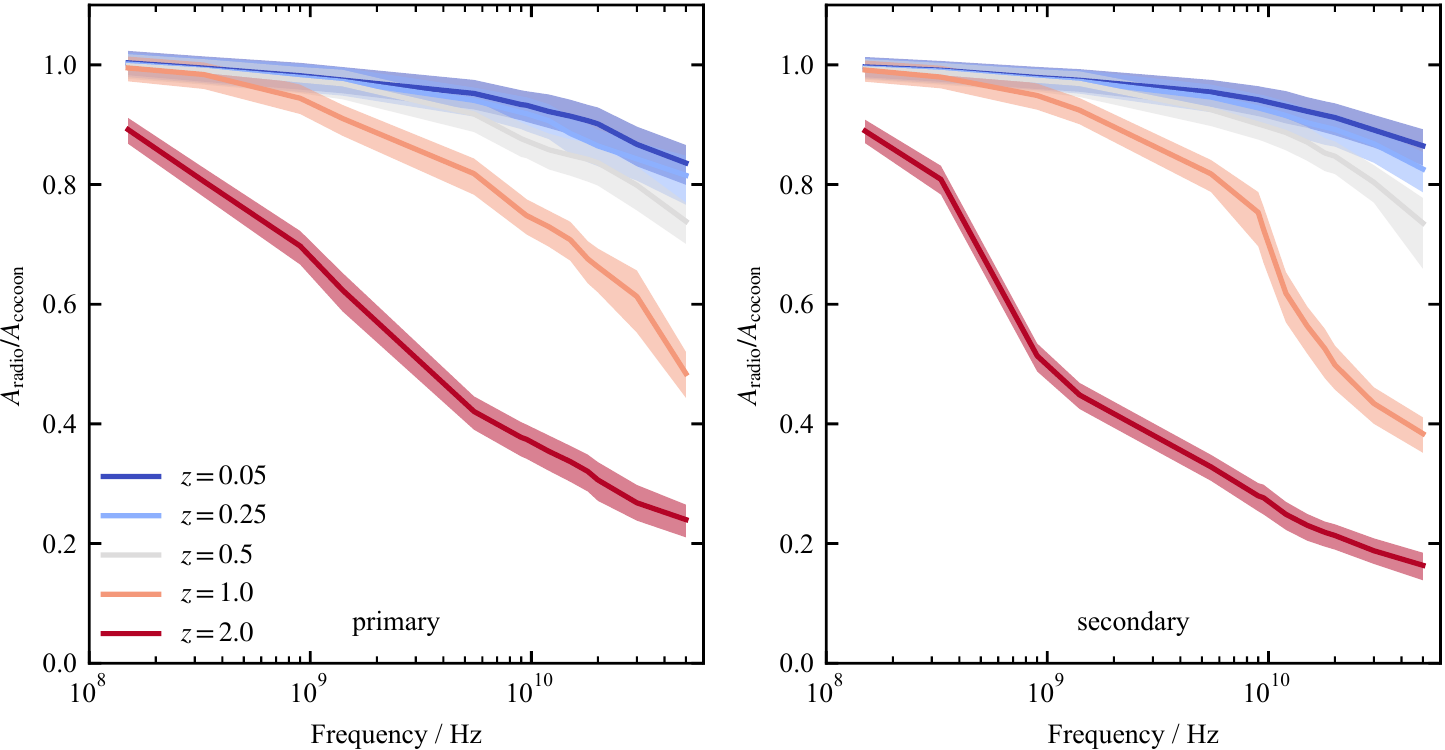}
    \caption{
      Ratio of observable radio lobe area to cocoon area, as a function of frequency for the $1\rcore$-offset simulation, at $t=32\myr$.
      Different colours show the detectability ratio at different redshifts.
      This ratio is shown for the primary jet in the left panel, and for the secondary jet in the right panel.
      The surface brightness sensitivity is taken to be lowest contour in \cref{fig:sr_radio-morphology}, $0.2\,\textrm{\textmu Jy arcsec}^{-2}$, corresponding to $0.05\,\textrm{mJy beam}^{-1}$ at $z=0.05$.
      A tracer cut-off of $10^{-7}$ is used to identify cocoon material.
      Shaded bands show the effect of changing the surface brightness sensitivity by a factor of 3.
    }
    \label{fig:sr_lobe_detectability}
  \end{figure*}

  \begin{figure*}
    \centering
    \includegraphics{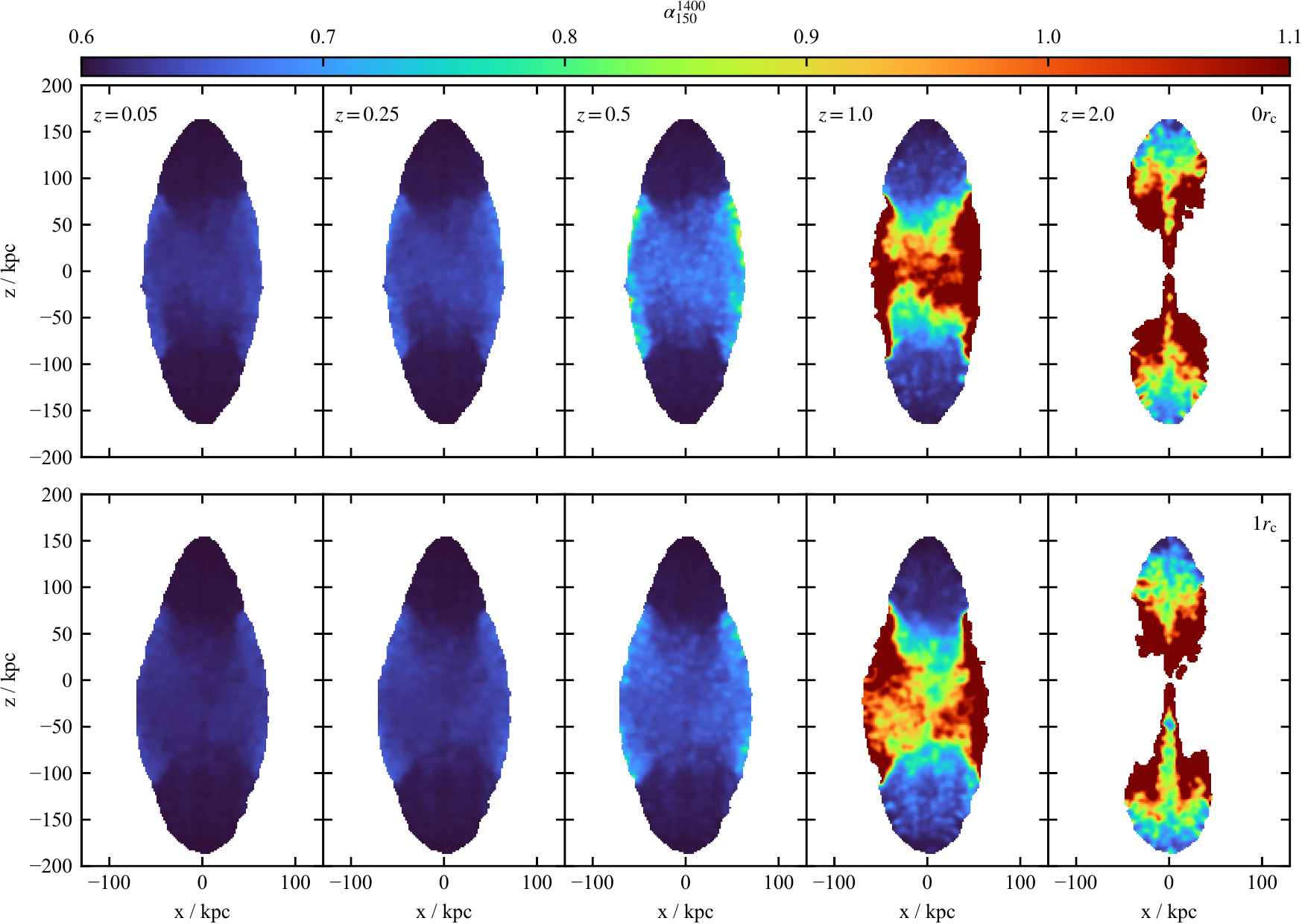}
    \caption{
      Plane of the sky spectral index maps between $150$ and $1400\,\textrm{MHz}$ at different redshifts, at $t=32\myr$.
      Physical pixel and beam size are as in \cref{fig:sr_radio-morphology}.
      The $0\rcore$-offset simulation is shown in the top row, while the $1\rcore$-offset simulation is shown in the bottom row.
    }
    \label{fig:sr_spectral_index_150_1400}
  \end{figure*}

  \begin{figure*}
    \centering
    \includegraphics{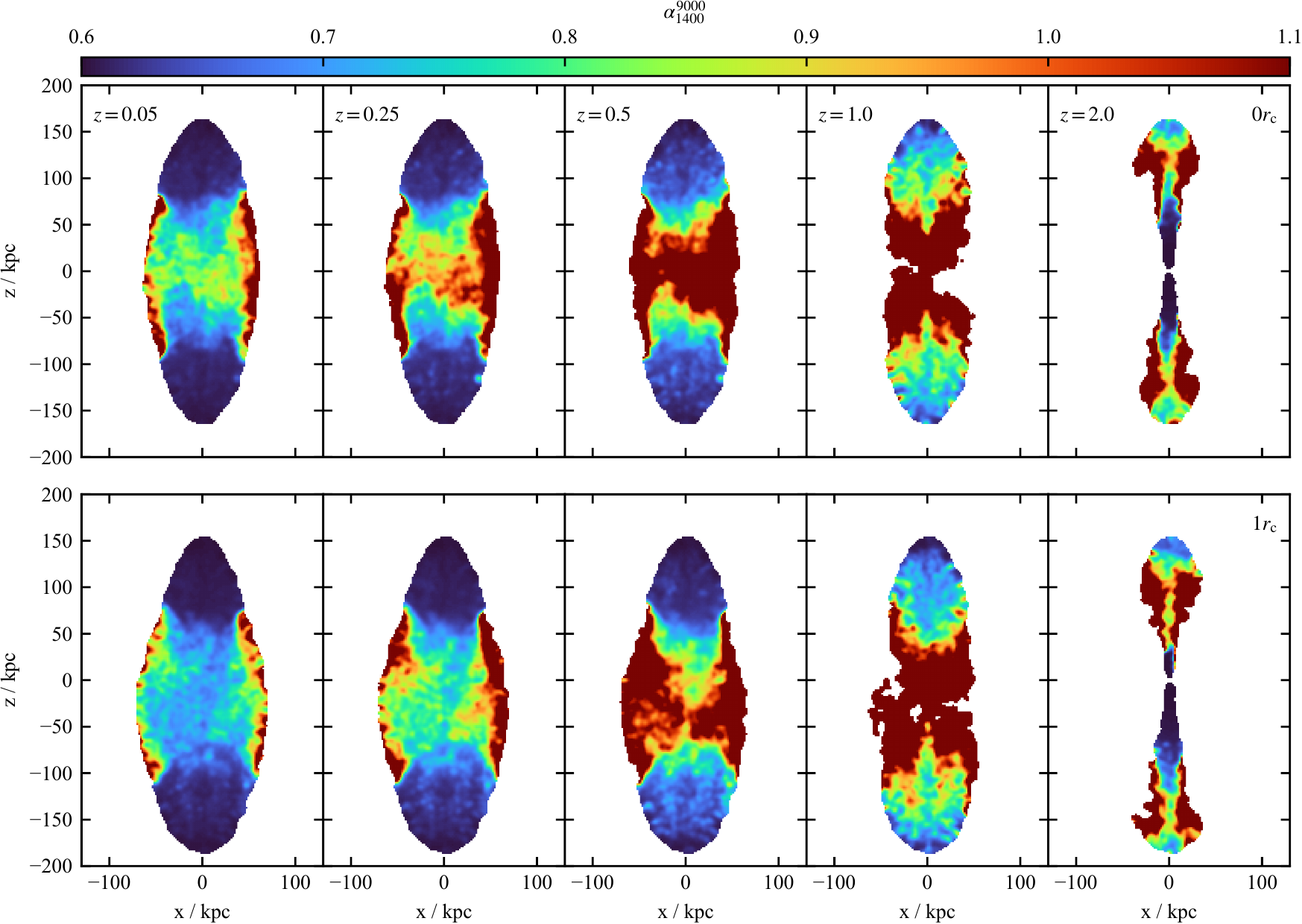}
    \caption{
      Same as \cref{fig:sr_spectral_index_150_1400}, but for spectral index between $1400$ and $9000\,\textrm{MHz}$.
    }
    \label{fig:sr_spectral_index_1400_9000}
  \end{figure*}

  \begin{figure*}
    \centering
    \includegraphics{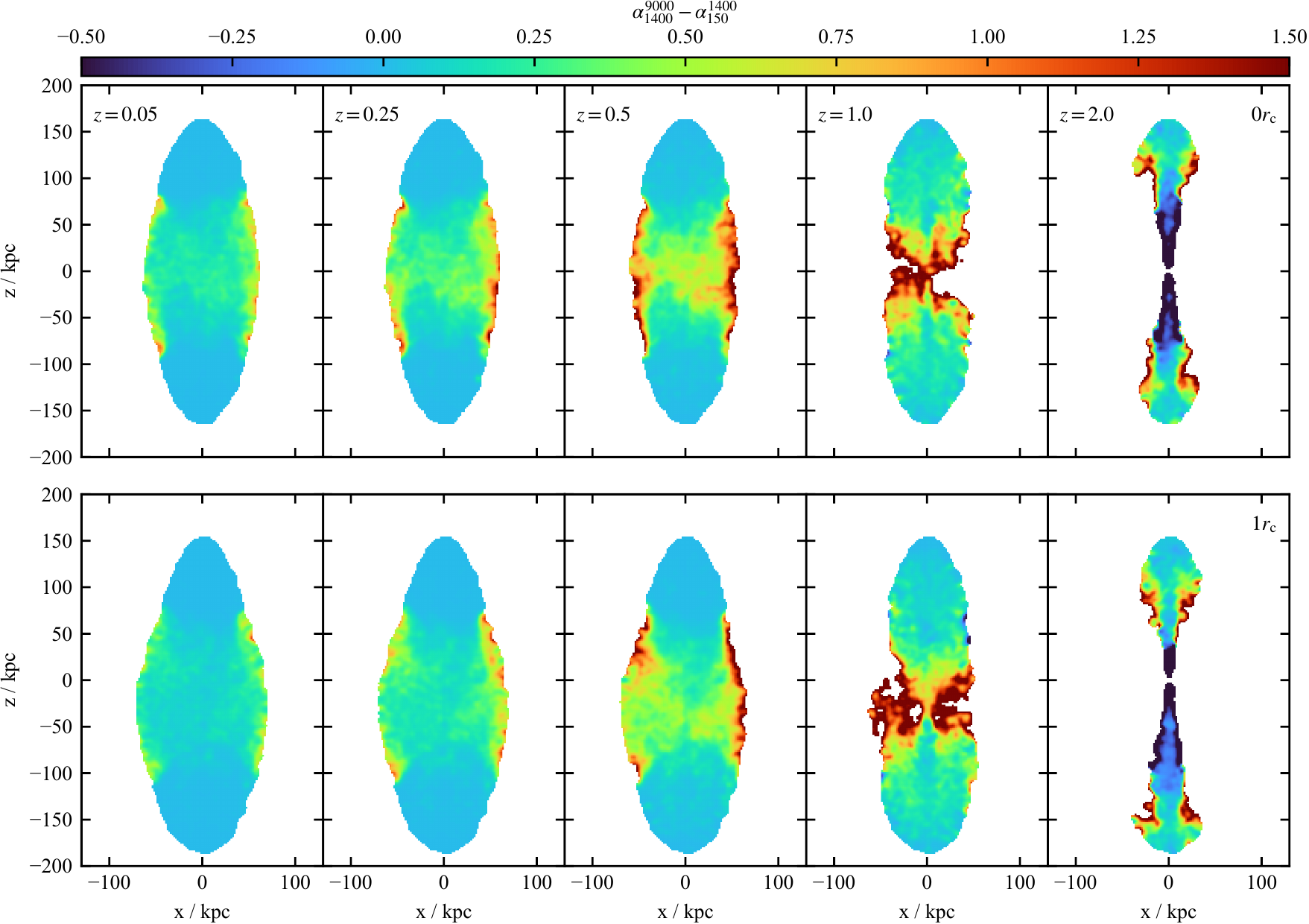}
    \caption{
      Plane of the sky spectral curvature $\left( \alpha_{1400}^{9000} - \alpha_{150}^{1400} \right)$ at $t=32\myr$.
      Rows and columns are as in \cref{fig:sr_spectral_index_150_1400}.
    }
    \label{fig:sr_spectral_curvature}
  \end{figure*}

  \begin{figure}
    \centering
    \includegraphics{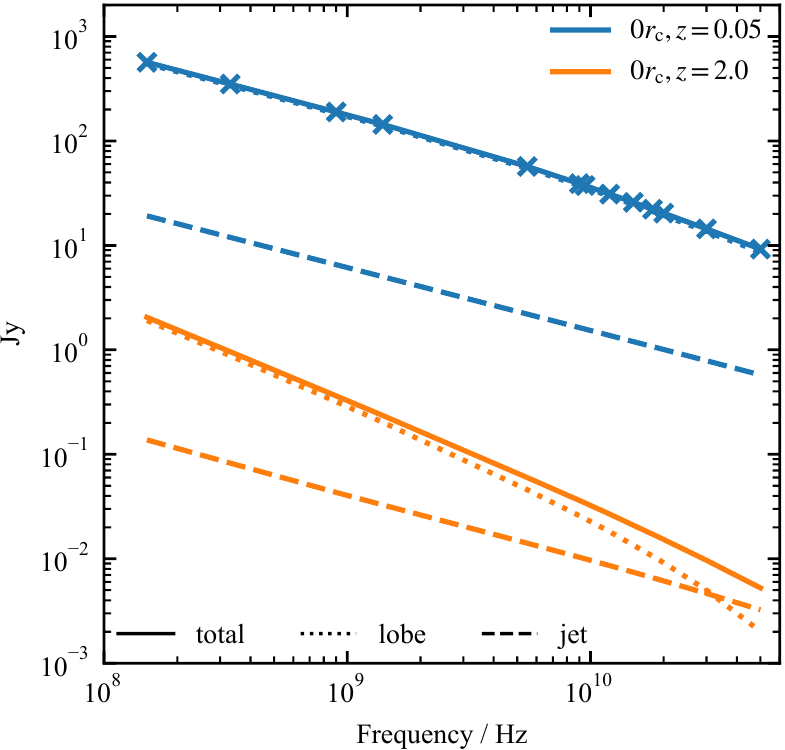}
    \caption{
      Integrated spectra for the $0\rcore$ simulation, at $z=0.05$ and $z=2.0$, for the same frequencies as \cref{fig:sr_lobe_detectability}, at $t=32\myr$.
      The integrated spectra for the $1\rcore$ simulation are identical, and hence not shown.
      Total source spectra are plotted as solid lines, while the separate lobe and jet components are plotted as the dotted and dashed lines respectively.
      Simulated observing frequencies are marked with crosses for the $0\rcore$, $z=0.05$ curve.
      Jet emission is determined using a particle velocity cut of $|\vec{v}| \ge 0.3c$.
    }
    \label{fig:sr_lobe_spectra}
  \end{figure}

  \begin{figure}
    \centering
    \includegraphics{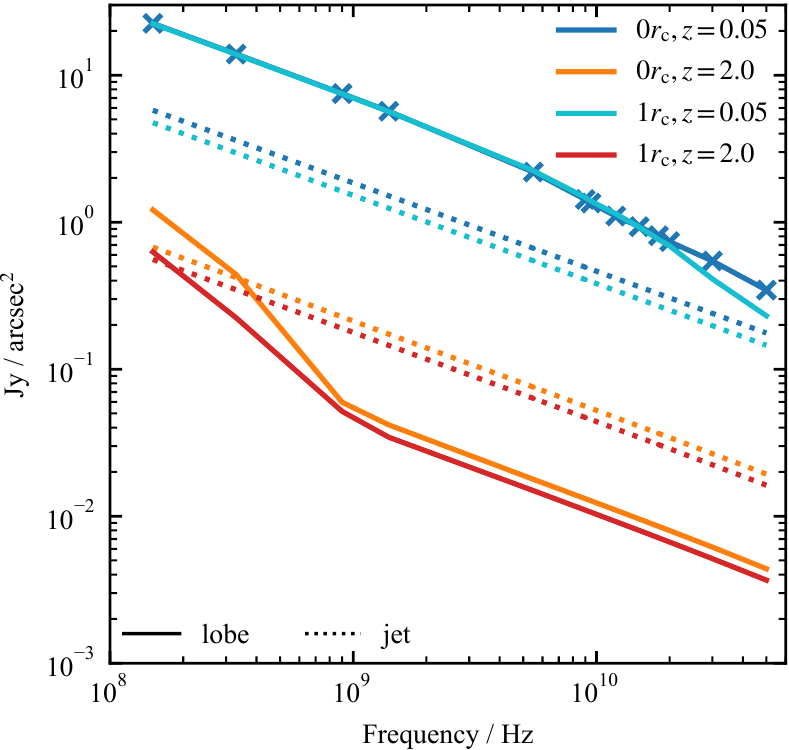}
    \caption{
      Local spectra for the southern jet at $t=32\myr$.
      The spectra are taken from a single pixel located at $(y,z)\approx(-60, 0)$; this pixel is marked on the surface brightness maps in \cref{fig:sr_radio-morphology} with a star.
      Jet and lobe contributions to the spectra are separated with the same velocity cut as in \cref{fig:sr_lobe_spectra}; here, lobe (jet) emission is plotted as the solid (dotted) line.
    }
    \label{fig:sr_local_spectra}
  \end{figure}

  \begin{figure*}
    \centering
    \includegraphics{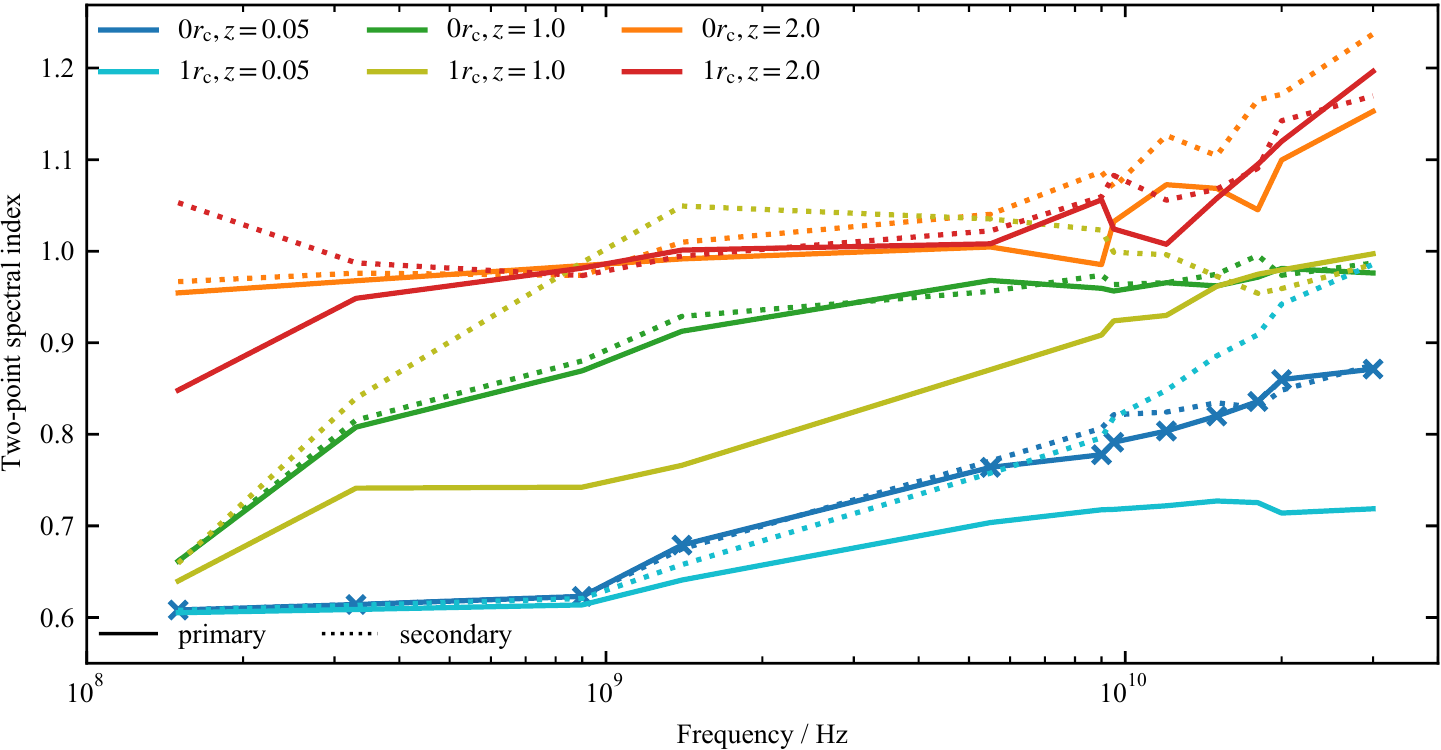}
    \caption{
      Integrated spectral index for the $0\rcore$ and $1\rcore$-offset simulations, at $z=0.05$, $z=1.0$, and $z=2.0$, plotted at the logarithmically spaced midpoint between $\nu_\textrm{low}$ and $\nu_\textrm{high}$ for the same frequencies as \cref{fig:sr_lobe_detectability}, at $t=32\myr$.
      The frequency midpoints are marked with crosses for the $0\rcore$, $z=0.05$ curve.
      The integrated spectral index for the primary lobe is plotted as the solid lines, while for the secondary lobe it is plotted as the dotted lines.
      The apparent asymmetry at $z=2.0$ for high frequencies in the $1\rcore$-offset simulation is a transient feature.
    }
    \label{fig:sr_two_point_spec_ind}
  \end{figure*}

  \subsection{Mapping between radio lobes and hydrodynamical structure}
  \label{sec:sr_radio_mapping}

    Comparing the radio morphology in \cref{fig:sr_radio-morphology} with the cocoon structure as shown by the density threshold in the left-hand panel of \cref{fig:sr_density-particle-distribution} shows that as the effect of losses increases (whether through increasing frequency or redshift), the observable radio lobes increasingly do not map to the underlying low-density jet cocoon.
    We quantify this in \cref{fig:sr_lobe_detectability}, plotting the ratio of observable radio lobe area to cocoon area as a function of both frequency and redshift, for the $1\rcore$-offset simulation.
    The observable radio lobe area is calculated assuming a sensitivity matching the lowest contour level shown in the surface brightness maps ($0.2\,\textrm{\textmu Jy arcsec}^{-2}$, \cref{fig:sr_radio-morphology}), while the cocoon area is calculated using a jet tracer cut-off, initially set to unity in the jet inlets only, to identify cocoon material.
    In this work, we classify the cocoon as cells with a tracer value $>10^{-7}$, noting that a tracer cut-off value of up to $10^{-3}$ gives similar results.
    The overall trends shown in \cref{fig:sr_lobe_detectability} with both frequency and redshift are largely independent of the tracer cut-off chosen.

    We find a significant evolution in the detectable fraction with redshift.
    At $1.4\,\textrm{GHz}$, the detectable fraction changes from $100\%$ at $z=0.05$ to between $40\%$ and $70\%$ at $z=2.0$.
    This result is a function of surface brightness sensitivity; the shaded bands in \cref{fig:sr_lobe_detectability} show the effect of changing the surface brightness sensitivity by a factor of $3$.

  \subsection{Spatially resolved losses}
  \label{sec:sr_spatially_resolved_losses}

    We now examine spatially resolved spectra for our simulated sources.
    The spectral index for a given pair of frequencies is calculated as $\alpha = - \log ( S_\textrm{high} / S_\textrm{low} ) / \log ( \nu_\textrm{high} / \nu_\textrm{low} )$.

    \subsubsection{Spectral index maps}
    \label{sec:sr_spec_ind_maps}

      \Cref{fig:sr_spectral_index_150_1400,fig:sr_spectral_index_1400_9000} show the low and high spectral index maps, $\alpha^{1400}_{150}$ ($\nu_\textrm{high}=1.4\,\textrm{GHz}$, $\nu_\textrm{low}=150\,\textrm{MHz}$) and $\alpha^{9000}_{1400}$ ($\nu_\textrm{high}=9\,\textrm{GHz}$, $\nu_\textrm{low}=1.4\,\textrm{GHz}$) respectively, for both simulations at $t=32\myr$.
      The spectral indices are shown for the same five redshifts as in \cref{fig:sr_radio-morphology}.

      The spectral index in the radio lobes agrees with expectations; for both the low and high spectral indices a constant spectral index is found in the hotspot region ($\alpha=0.6$, reflecting the adopted electron energy injection index $s=2.2$), indicative of a significant population of recently accelerated electrons.
      The spectral index steepens away from the lobe tips, as electrons accelerated there flow back into the lobes, losing energy and mixing with older electron populations.
      This increase of the spectral index away from the lobe tips is more pronounced at both higher redshifts, where inverse-Compton losses are greater, and for higher frequencies, where synchrotron losses are greater (compare \cref{fig:sr_spectral_index_150_1400} and \cref{fig:sr_spectral_index_1400_9000}).

      When significant losses are present (whether due to high frequencies or high redshifts), the jet is prominent in the spectral index maps.
      This occurs because the jet in our simulations is shocked early after recollimation, and also because of the rapid energy loss experienced by the emitting electrons population after acceleration, resulting in only the most recently accelerated electrons emitting at the observed frequencies.
      These electrons are typically found near the jet and hotspots.

    \subsubsection{Spectral curvature maps}
    \label{sec:sr_spec_curvature_maps}

      In \cref{fig:sr_spectral_curvature} we show the spectral curvature of the radio lobes, defined as $\Delta\alpha = \alpha^{9000}_{1400} - \alpha^{1400}_{150}$, for both simulations at $t=32\myr$.
      A large $\Delta\alpha$ means that the local spectrum is steeper at higher frequencies, while zero means it is a power-law with a constant slope.
      The spectral curvature shows significant spatial evolution with frequency.
      The lobe tips and immediate surrounding regions have a spectral curvature of zero, indicative of a population of young electrons.
      At low redshifts, this power-law region extends almost back to the jet core; however, this is not the case at higher redshifts.
      We find significant steepening of the spectra due to losses in the equatorial regions, shown by increasing spectral curvature, indicative of both a mixed and ageing population of electrons.
      Due to the jet instabilities, complex shock histories are present in the downstream turbulent flow, as shown in \cref{fig:sr_density-particle-distribution}.
      The local shock structure in this unstable region consists of many local shocks; some electron packets are shocked very recently, while others are shocked early and thereafter only lose energy without any subsequent shocks to boost their energy distribution.
      This combination of a freshly shocked population with a cooling population in the same region results in a complex spectral curvature map.

    \subsubsection{Integrated spectra}
    \label{sec:sr_integrated_spec}
    
      Jet emission begins to dominate over lobe emission at high redshifts; this leads to the negative spectral curvature present along the jet for redshifts $z=1.0$ and $z=2.0$.
      We explore this effect in more detail by splitting the emission into jet and lobe components.
      In \cref{fig:sr_lobe_spectra} the total, lobe, and jet integrated spectra are plotted--the solid, dotted, and dashed lines respectively--for both simulations at $z=0.05$ and $z=2.0$.
      Particles with velocities greater (less) than $0.3 c$ are classified as jet (lobe) material.
      The jet spectra have the constant slope expected from a young electron population, while some curvature is evident in the lobe spectra.
      At low redshifts, emission from the radio lobes dominates the spectrum at all frequencies.
      However, at high redshifts, the ageing lobes experience strong inverse-Compton losses and the lobe spectrum becomes comparable to that of the jet.
      Higher frequencies accentuate this effect, and the total spectrum is flatter in this region than at low frequencies.

    \subsubsection{Local spectra}
    \label{sec:sr_local_spec}

      \cref{fig:sr_local_spectra} shows the local lobe and jet spectra (solid and dotted lines respectively) for a specific pixel located in the southern jet, marked by the white star in the right-most panels of \cref{fig:sr_radio-morphology}.
      The local jet spectra exhibit no curvature, consistent with the integrated spectra.
      However, the local lobe spectra steepen at higher frequencies and low redshifts while still dominating over the jet component.
      At high redshifts, contributions from the jet dominate the spectra for all but the two lowest frequencies studied, due to the significant curvature in the lobe spectra.
      These local spectra highlight the different electron populations responsible for producing the integrated spectra, and explain the negative spectral curvature found at the highest frequencies at higher redshifts.

    \subsubsection{Integrated spectral index}
    \label{sec:sr_integrated_spec_ind}

      In \cref{fig:sr_two_point_spec_ind} we plot the integrated spectral index as a function of frequency for each simulation, at redshifts $z=0.05$, $z=1.0$, and $z=2.0$.
      The spectral index $\alpha^{\nu_\textrm{high}}_{\nu_\textrm{low}}$ is computed between adjacent frequency pairs, for the $11$ observing frequencies used in \cref{fig:sr_lobe_detectability}.
      We find significant evolution of spectral index with frequency.
      The spectrum steepens with frequency for all lobes at $z=0.05$ as radiative losses become more important, and significant differences are observed in the integrated spectral index at $z=0.05$ between the two radio lobes of the $1\rcore$-offset simulation: the lobe expanding into a falling density profile has a steeper spectrum than the lobe expanding into a rising density profile, indicative of greater losses.
      From this, we conclude that the environment plays a significant role in the losses across a radio lobe and that the steeper spectrum in the secondary lobe is caused by the faster lobe expansion and consequently larger volume.
      This causes electrons to experience greater adiabatic losses after they are accelerated, requiring larger accelerated Lorentz factors $\gamma_\textrm{acc}$ to emit at the present time with $\gamma(\nu)$ than if the adiabatic losses were smaller.
      Higher Lorentz factors at earlier times also result in greater synchrotron and inverse-Compton losses, which are proportional to $\gamma^2$.

      At higher redshifts this asymmetry is still present, but it is systematically shifted towards lower frequencies with increasing redshift due to a combination of increased inverse-Compton losses and the emitting-to-observed frequency shift.
      This has an observational impact: the spectral index asymmetry is highlighted at different frequencies for different redshifts, so using these differences as environmental probes requires an understanding of the relevant frequency range.

\section{Discussion}
\label{sec:sr_discussion}

  \subsection{Reproducing observed radio properties}
  \label{sec:sr_praise_verification}

    In \cref{sec:sr_validation,sec:sr_results}, we have demonstrated that the method used in this paper to model non-thermal emission from radio sources produces results that are consistent with expectations.
    The numerical hydrodynamic model for jet evolution tracks electron population mixing within the cocoon with greater accuracy than semi-analytic models, and produces more complex and varied electron populations and histories.
    Spatially resolved shock acceleration also affects the synchrotron signatures of emitting electrons within a given region of the cocoon; accurate description of this process requires numerical hydrodynamics.

    In \cref{fig:sr_lobe_detectability} we find that the observed radio lobe area decreases as electron losses become more pronounced at both higher frequencies and redshifts.
    This phenomenon is consistent with expectations, and is reflected in observations \citep[e.g., the pinched lobes of Cygnus A at higher frequencies; ][]{Carilli1991}.
    The expected steepening of the spectral index away from the lobe tips is also reproduced; this is evident from observations, and consistent with our earlier work \citepalias{Turner2018a}.
    At higher redshifts, our simulated radio sources have steeper spectral indices in agreement with observations \citep{Morabito2018}.

  \subsection{Effect of environment on radio observables}
  \label{sec:sr_effect_of_environment}

    In \citetalias{Yates-Jones2021} we showed that environment plays a significant role in radio source evolution and dynamics.
    In this work, we show that environment also plays an important role in determining the radio observables, by affecting electron spectra.
    We are sampling three different isothermal environments with the two simulations presented here: both jets in the $0\rcore$-offset simulation are expanding into identical environments, and hence act as the control case; meanwhile, the primary jet in the $1\rcore$-offset simulation is expanding into a rising density profile, while the secondary jet is expanding into a falling density profile.
    We find that the primary lobe for the $1\rcore$-offset simulation has brighter regions at the lobe tips at all redshifts and frequencies, compared to the secondary lobe.
    Small-scale differences in lobe morphology are observed between the primary and secondary radio lobes.
    However, the evolutionary tracks through the size-luminosity diagram show little dependence on environment for a large source size.
    The different lobe morphologies are reflected in the different observable area ratios shown in \cref{fig:sr_lobe_detectability} (comparing left and right panels).
    The secondary lobe has a narrower, more pinched lobe at all redshifts for frequencies above $\sim 10^{9}\,\textrm{Hz}$.

    It is the integrated spectral index that most clearly highlights the role environment plays in determining the loss process of emitting electrons.
    The primary lobe of the $1\rcore$-offset simulation has a significantly flatter spectral index than the secondary lobe across the observing frequencies, with $\Delta\alpha \sim 0.1$ at GHz frequencies; the primary lobe spectra are also flatter than those of the lobes in the $0\rcore$-offset simulation.
    Meanwhile, the secondary lobe in an asymmetric environment shows significant steepening across all frequencies.
    This difference is due to the different cocoon dynamics between the two lobes, as the primary expands into a rising pressure profile, while the secondary expands into a falling pressure profile.
    The different dynamics lead to a different magnitude of losses in the lobes, producing the observed result.

  \subsection{Comparison to analytic models and in-situ electron energy evolution}
  \label{sec:sr_pros_and_cons}

    The strength of PRAiSE lies in applying the analytic iterative loss processes of \citetalias{Turner2018a} to purely hydrodynamic simulations.
    This makes use of fluid (back)flow (rather than dynamical models) for the electron packet positions (using Lagrangian tracer particles advected with the fluid), and numerical pressure histories as encountered by the tracer particles.
    PRAiSE can be applied to model any radio source since the underlying dynamics are calculated with numerical simulations; existing analytic models for radio lobes \citep{Turner2018a,Hardcastle2018} are restricted to sources with dynamics that can be described analytically, such as simple FR~I and FR~II morphologies.
    Even for simple FR~II morphologies, differences in dynamics between analytic and numerical models exist.
    For example, \citetalias{Turner2018a} distributes injected energy self-similarly to drive the jet cocoon evolution.
    In numerical simulations, where this assumption is removed \citep[e.g.,][]{Hardcastle2013}, we find that energy is initially preferentially deposited at the hotspots, leading to a swift initial expansion, i.e. a ``jet breakout'' phase

    The core assumptions of \citetalias{Turner2018a} still apply to our post-processing, most importantly that particle acceleration occurs only at the site of strong shocks (although these are no longer confined to be at the hotspots), and an identical population of electrons is injected at each shock; the injection index is not dependent on shock properties.
    In this way, it is possible to gauge the numerical work against the results from the analytical models.
    In future work, we plan to relax these assumptions and, for example, couple the injection index to the shock strength and evolve the magnetic field dynamically with the simulation.
    Recent work \citep{Matthews2019,Bell2019} has highlighted the importance of weak shocks and turbulence in backflows to accelerating ultrahigh energy cosmic rays, which are required for a complete treatment of radio lobe emissivity.
    This should then be fully comparable to more complex models of non-thermal emission \citep{Mendygral2012,Vaidya2018,Mukherjee2020}, which evolve the electron population in-situ according to the strength of shocks on the grid, but with the added advantage that the source can be placed at different redshifts and the particle acceleration physics can be changed in post-processing without the need to re-run the simulation.

    Finally, we reiterate that while the PRAiSE method can use magnetic field energy densities directly, in this work we have used purely hydrodynamic quantities, assuming a constant departure from equipartition.
    This has the effect of smoothing out small fluctuations due to localised changes in the magnetic field, which has been shown to occur in radio lobes with magnetohydrodynamic simulations \citep[e.g.,][]{Gaibler2009, Hardcastle2014}.

\section{Conclusions}

  In this paper, we have presented the PRAiSE framework for resolved spectral evolution in radio sources.
  We use PRAiSE to calculate the synthetic synchrotron emissivity for hydrodynamic simulations of radio jets, incorporating adiabatic and radiative loss processes through the use of Lagrangian tracer particles that each carry an electron packet.
  We combine this with an effective tessellation of the computational domain and show that all radio emitting structures are well sampled.
  The method addresses loss processes and spatially resolved spectral ageing very well, and we demonstrate that emission and spectral index maps at a variety of frequencies can be produced.
  The method promises huge savings in computational resources, as different source redshifts and shock physics can be explored in post-processing with the same simulation.

  We reproduce the observed strong dependence of radio source spectral index with redshift due to inverse-Compton losses \citep{Morabito2018}.
  Additionally, we find a dependence of the spectral index on environment.
  Radio sources in denser environments have flatter spectral indices; this difference is particularly pronounced between the two lobes of our simulation in an asymmetric environment.
  In addition, we find the asymmetry in spectral index between two lobes to have a systematic dependence on redshift and observing frequency.

  Our jets disrupt before reaching the tip of the lobe due to a low internal Mach number.
  While this has been suggested to be able to explain an FR~I morphology, our results do not support this suggestion, as the most prominent radio emission site remains near the tip of the lobes despite the early disruption of the jet.
  Finally, we note that the observed radio structure does not map to underlying morphology, and it is challenging to infer the extent of jet feedback from radio observations alone; we defer to a future paper a detailed discussion of the mapping between radio observables and feedback.

\section*{Acknowledgements}

  We thank an anonymous referee for their useful comments.
  PYJ thanks the University of Tasmania for an Australian Postgraduate Award, the ARC Centre of Excellence for All Sky Astrophysics in 3 Dimensions for a stipend, and both the University of Tasmania and the Astronomical Society of Australia for their international travel support.
  SS thanks the Australian Government for an Endeavour Fellowship 6719\_2018.
  PYJ and SS thank the Centre for Astrophysics Research at the University of Hertfordshire for their hospitality.\\
  \noindent This work was supported by resources awarded under Astronomy Australia Ltd’s ASTAC merit allocation scheme, with computational resources provided by the National Computational Infrastructure (NCI), which is supported by the Australian Government.
  We would like to acknowledge the use of the high performance computing facilities provided by Digital Research Services, IT Services at the University of Tasmania.
  We acknowledge the work and support of the developers providing the following Python packages: Astropy \citep{AstropyCollaboration2018,AstropyCollaboration2013}, JupyterLab \citep{Jupyter}, Matplotlib \citep{Matplotlib}, Numba \citep{Numba}, NumPy \citep{NumPy}, and SciPy \citep{SciPy}.

\section*{Data Availability}

  The data underlying this article will be shared on reasonable request to the corresponding author.



  \bibliographystyle{mnras}
  \bibliography{synth-radio-local.bib}




  \bsp	
  \label{lastpage}
\end{document}